\documentstyle[pre,aps,epsfig,twocolumn]{revtex}
\begin{document}
\draft
\title{\large \bf Exploring phase space localization of chaotic eigenstates via
parametric variation}
\author{Nicholas R. Cerruti, Arul Lakshminarayan,\footnote{Permanent address: 
Physical Research Laboratory, Navrangpura, Ahmedabad 380 009, India.}
Julie H. Lefebvre,\footnote{Current address:  Defence Research Establishment Ottawa, 
Ottawa, Ontario, K1A 0Z4, Canada}
and Steven Tomsovic}
\address{Department of Physics, Washington State University, Pullman, WA
99164-2814, USA}
\address{$^{*}$Permanent address: 
Physical Research Laboratory, Navrangpura, Ahmedabad 380 009, India.}
\address{$^{\dagger}$Current address:  Defence Research Establishment Ottawa, 
Ottawa, Ontario, K1A 0Z4, Canada}
\date{\today}
\maketitle

\begin{abstract}

In a previous Letter [Phys.~Rev.~Lett.~{\bf 77}, 4158 (1996)], a new
correlation measure was introduced that sensitively probes phase space
localization properties of eigenstates.  It is based on a system's response
to varying an external parameter.  The measure correlates level
velocities with overlap intensities between the eigenstates and some localized
state of interest.  Random matrix theory predicts the absence of such 
correlations in chaotic systems whereas in the stadium billiard, a paradigm of 
chaos, strong correlations were observed.  Here, we develop further the 
theoretical basis of that work, extend the stadium results to 
the full phase space, study the $\hbar$-dependence, and
demonstrate the agreement between this measure and a semiclassical theory based
on homoclinic orbits. 
\end{abstract}


\begin{section}{Introduction}
\label{intro}

The two general motivations for our investigation are 
understanding better the nature of eigenstates of bounded quantum systems
possessing `simple' classical analogs, and exploring new features of such
systems' behavior as a system parameter is smoothly varied.  Simple in this
context refers to few degrees of freedom and a compact Hamiltonian.  
Nevertheless, the classical dynamics may display a rich variety of
features from regular to strongly chaotic motion.  We focus on the
strongly chaotic limit for which semiclassical quantization of
individual chaotic eigenstates does not hold, and the
correspondence principle is less well understood~\cite{einstein}.  
Even though there has been some recent progress~\cite{th3}, it turns
out that with a detailed understanding of chaotic systems a 
statistical theory provides a well-developed, alternative approach
to these difficulties.  Twenty years ago, Berry~\cite{berry} conjectured
and Voros~\cite{voros} discussed that in this case as $\hbar \rightarrow 0$
the eigenstates should respect the ergodic hypothesis in phase space,
$\delta(E-H({\bf p},{\bf q}))$, as it applies to wavefunctions.  In essence,
the eigenmodes should appear as Gaussian random wavefunctions locally in
configuration space with their wavevector constrained by the ergodic measure of
the energy surface.  Discussion of the properties of random waves and recent
supporting numerical evidence can be found in refs.~\cite{hog,srednicki}.  

The second general motivation relates to a long recognized class of
problems, i.e.~a system's response to parametric variation.  Our interest here
is restricted to external, controllable parameters such as electro-magnetic
fields, temperatures, applied stresses, changing boundary conditions, etc...,
through whose variation one can extract new information about a system not
available by other means.  A multitude of examples can be found in the
literature~\cite{exampleparam}.  A recent concern has been universalities in the
response of chaotic or disordered systems and statistical
approaches to measuring the response~\cite{marcus}.  Universal parametric
correlations have been derived via field theoretic or random matrix methods for
quantities involving level slopes (loosely termed velocities in this paper), 
level curvatures, and eigenfunction amplitudes~\cite{altsimon,alhassid}.  
In contrast, our motivation is not the universal features per se for 
they cannot tell us anything specific about the system other than it is, 
in fact, chaotic and/or symmetry is present.  Rather we are interested 
in what system specific information can be extracted in the case 
that the system's response deviates from universal statistical laws. 
The specific application discussed in this paper shows how one can decipher phase
space localization features of the eigenstates.  The theory naturally divides
into a two-step process.  One must first understand any implied limiting
universal response of chaotic systems.  Next, one must develop a theory
which gives a correct interpretation of any deviations seen from the universal
response.  The necessarily close interplay between theory and observation
required to deduce new information forms part of the attractiveness of
investigating parametric response.  

Taking up the first step of understanding universal response, an expected but
rarely discussed property is the independence of eigenvalue and eigenfunction
fluctuation measures~\cite{mehta} which is found in the random matrix theories
anticipated to describe the statistical properties of quantum systems with
chaotic classical analogs~\cite{bgs,aaa}.  Coupled with Berry's
conjecture mentioned above, these properties imply a `democratic' response
to parametric variation for an ergodically behaving quantum system.  The
perturbation connects one state to all other states locally with equal
probability.  The variation of any one eigenstate or eigenvalue over a large
enough parameter range will be statistically equivalent to their respective
neighboring states or levels.  

In a pioneering work on the ergodic hypothesis using the stadium billiard, 
now a paradigm of chaos studies, McDonald
noticed larger than average intensities of the eigenstates in certain 
regions~\cite{mcdonald}.  In his thesis he
states that ``a small class of modes (bouncing ball, whispering gallery, etc.) 
seem to correspond naively to a definite set of `special' ray orbits.''  
Heller initiated a theory concerning these large intensities when 
he modified the random wavefunction picture with his prediction and 
numerical observations of eigenstate scarring~\cite{heller}.  He derived a 
criterion for eigenstate intensity in excess of the ergodic predictions along
the shorter, less unstable periodic orbits.  Scarring is thus one possible
phase space `localization' property of a chaotic eigenstate.  Other
possibilities result from time scales not related directly to the Lyapunov
instability such as transport barriers in the form of broken
separatrices~\cite{btu}, and cantori~\cite{percival,geisel}, or diffusive
motion~\cite{fishgremprang}.  In the context of this paper, 
we take localization to
mean some deviation from the ergodic expectation beyond the inherent quantum
fluctuations, and it creates the possibility of a non-democratic response to
parametric variation.  A perturbation could preferentially connect certain
states or classes of states, thus leading to additional short-range avoided
crossings or like level movements within a particular class, etc. 

Debate ensued Heller's work on eigenstate scarring, in part, because of the
difficulty in quantitatively characterizing and predicting its extent in either
a particular eigenstate or even collective groups of eigenstates.  Judging from
the earlier literature, it was easier to graph eigenstates in order to see the 
scarring by eye than define precisely what it means or what its physical significance
is.  Furthermore, he linearized the semiclassical theory which was
insufficient for a full description of scarring.  We remark that
recent work suggests the opposite, i.e.~the 
linearized theory is sufficient assuming $\hbar$ is
smaller than some system specific value which is `small enough'~\cite{hk}. 
However, many of the experimental and numerical investigations are far from this
regime and the nonlinear dynamical contributions are essential for understanding
most of the work being done.  The theory incorporating nonlinear dynamical
contributions~\cite{th3,th1} was developed much later than Heller's
introduction of scarring.  It is based on
heteroclinic orbit expansions for wave packet propagation and strength
functions.  Ahead, we make extensive use of these forms to derive a
semiclassical theory applicable to problems involving parametric variation.  

In a previous Letter~\cite{tomsovic}, one of us (ST) introduced a measure that
very sensitively probes phase space localization for systems having
continuously tunable parameters in their Hamiltonians.  It correlates level
motions under perturbation with overlap intensities between eigenstates and
optimally localized wave packet states.  The basic idea is that the wave
packet overlap intensities select eigenstates that potentially have excess
support in the neighborhood of the phase point at the wave packet's position 
and momentum centroids.  The perturbation will push these levels somewhat in 
the same direction depending on how it is distorting the energy surface near 
that particular phase point.  If the level velocities associated with those 
states have similar enough values, then significant non-zero correlations will
result that reveal the localization.  The measure can be used in a forward or
reverse direction.  If phase space localization is present in a system of
interest, then it predicts experimentally verifiable manifestations of that
localization.  Conversely, one can first experimentally determine the level
velocity - overlap intensity measure in that system for the purpose of
inferring the existence and extent of localization.  

Our purpose in this paper is to give a complete account of that Letter, develop
further the semiclassical theory, and explore the full phase space and
$\hbar$ behavior of the stadium billiard, a continuous time system.  In a
companion paper immediately following this one, we give the theory for
quantized maps (discretized time)~\cite{lakshminarayan2}.  The
next section introduces strength functions and a new class of correlation
coefficients.  Section~\ref{rmt} utilizes ergodicity and random wave properties
to motivate the introduction of random matrix ensembles.  The ensembles
describe the statistical properties of chaotic systems in the
$\hbar\rightarrow 0$ limit.  The correlation measures vanish for these
ensembles indicating the absence of localization and universal
response to perturbation (i.e.~parameter variation). 
Section~\ref{semi_dynamics} gives the semiclassical theories of level velocities,
strength functions, and overlap intensity-level velocity correlation coefficients.
We finish with a full treatment of the stadium billiard and concluding remarks. 

\end{section}

\begin{section}{Preliminaries}
\label{prelim}

Consider a quantum system governed by a smoothly parameter-dependent
Hamiltonian, $\hat H (\lambda)$ with classical analog $H({\bf p},{\bf q};
\lambda)$.  We suppose that the dynamics are chaotic for all values of the
$\lambda$ range of interest, and suppose the absence of symmetry breaking.  
Then the expectation is that all statistical properties are stationary with 
respect to $\lambda$.  Without loss of generality, we also assume the phase 
space volume of the energy surface is constant as a function of $\lambda$.  
This ensures that the eigenvalues do not collectively drift in some direction 
in energy, but rather wander locally.  We use the same strength function Heller
employed in his prediction of scarring~\cite{heller} except slightly
generalized to include parametric behavior;
\begin{eqnarray}
  \label{strengthfunction}
  S_\alpha(E,\lambda)&=&{1\over 2\pi\hbar}\int^\infty_{-\infty} \ {\rm d}t\
    e^{iEt/\hbar}\langle \alpha|e^{-i\hat H (\lambda)t/\hbar}
    |\alpha\rangle \nonumber \\
  &=& Tr[\hat p_\alpha \delta (E - \hat H (\lambda))] \nonumber \\
  &=& \sum_n p_{\alpha n}(\lambda) \delta (E - E_n(\lambda)); \\ 
  p_{\alpha n}(\lambda) &=& |\langle \alpha |E_n(\lambda)\rangle|^2 
    \nonumber
\end{eqnarray}
where $\hat p_\alpha = |\alpha\rangle\langle\alpha|$.  $S_\alpha(E,\lambda)$ 
is the Fourier transform of the autocorrelation function of a special
initial state $|\alpha\rangle$ of interest.  
Ahead $\overline{S}_\alpha(E,\lambda)$ will
denote the smooth part resulting from the Fourier transform of just the
extremely rapid initial decay due to the shortest time scale of the dynamics
(zero-length trajectories).  We will take $|\alpha\rangle$ to be a Gaussian
wave packet because of its ability to probe ``quantum phase space,'' 
but other choices are possible.  Say momentum space localization were the main
interest, the natural choice would be a momentum eigenstate. $|\alpha\rangle$ 
can be associated with a phase space image $\rho_{\alpha}({\bf p}, {\bf q})$ 
of Gaussian functional form using Wigner transforms or related techniques.
$\rho_{\alpha}({\bf p}, {\bf q})$ turns out to be positive definite and maximally
localized in  phase space, i.e.~it occupies a volume of $h^{d}$.   

For a fixed value of the parameter, an example strength function is shown in
Fig.~(\ref{strength}).  If the wave packet is centered somewhere on a short
periodic orbit, large amplitudes necessarily indicate significant wave
intensity all along the orbit as seen in the inset eigenstates.  This behavior
cannot, a priori, be stated to be obviously in violation of the
quantum statistical fluctuation laws even if it appears so.  That remains to be
determined.  With the inclusion of parametric variation, the eigenvalues of a
chaotic system are supposed to move along smoothly varying curves of the type
shown in the upper square of Fig.~(\ref{level_velocities}).  Many of the
previous studies of parametric variation focussed on the properties of such
level curves.  A great deal is known about the distribution of level
velocities~\cite{altshuler,mehlig}, the decay of correlations in parametric
statistics~\cite{alhassid,simons}, the distribution of level
curvatures~\cite{gaspard,delande,vonoppen}, and the statistics of the 
occurrences of avoided crossings~\cite{delande2,wilkinson}.

\begin{figure}
  \epsfig{file=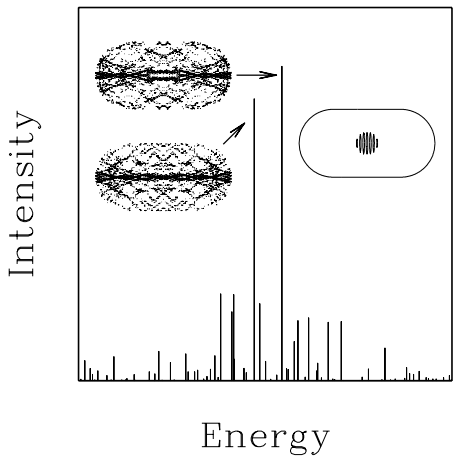, width = 7.4cm}
  \caption{Strength function for the stadium billiard.  The Gaussian wave packet
  is centered in the stadium with momentum directed towards the end cap.  The 
  large intensities are where the scarring occurs.}
  \label{strength}
\end{figure}

We now superpose the strength function overlap intensity information
on Fig.~(\ref{level_velocities}) in the lower square as vertical lines centered
on the levels; the lengths are scaled by the intensities (3-D versions of
this figure turned out not to be very helpful).  By considering the full
strength function and not just the level curves (i.e.~density of states), the
eigenstate properties can be more directly probed.  A new class of statistical
measures can be defined that cross correlate intensities with levels.  The
most evident examples are the four correlation coefficients 
involving both level curves and eigenstate amplitudes that
can be defined from the following  quantities: (i) the level velocities, 
$\partial E_n(\lambda) /\partial \lambda$, (ii) level curvatures, 
$\partial^2 E_n(\lambda) /\partial \lambda^2$,  
(iii) overlaps, $p_{\alpha n}$, and (iv) overlap changes, 
$\partial p_{\alpha n} / \partial \lambda$.  The most important 
is the overlap intensity-level velocity correlation coefficient,
${\cal C}_\alpha (\lambda)$, which is defined as
\begin{equation}
  \label{correlator}
  {\cal C}_\alpha (\lambda)= {\left\langle p_{\alpha n} {\partial
  E_n(\lambda) \over \partial \lambda}\right\rangle_E \over
  \sigma_\alpha \sigma_E}
\end{equation}
\noindent where $\sigma_\alpha^2$ and $\sigma_E^2$ are the local variances of the 
overlaps and level velocities, respectively.  The brackets denote a local
energy average in the neighborhood of $E$.  It weights most the level 
velocities whose associated eigenstates possibly
share common localization characteristics and measures the tendency of
these levels to move in a common direction.  In this expression,
the phase space volume remains constant so that the level velocities are 
zero-centered (otherwise the mean must be subtracted), and ${\cal C}_\alpha
(\lambda)$ is rescaled to a unitless  quantity with unit variance making it a
true correlation coefficient.  The set of states included in the local
energy averaging can be left  flexible except for a few constraints.  Only
energies where $\overline{S}_{\alpha}(E, \lambda)$ is roughly constant can be
used or some intensity unfolding must be applied.  Also, the energy range must be
small so that the classical dynamics are essentially the same throughout the
range, but it must also be broad enough to include several eigenstates.  
\begin{figure}
  \epsfig{file=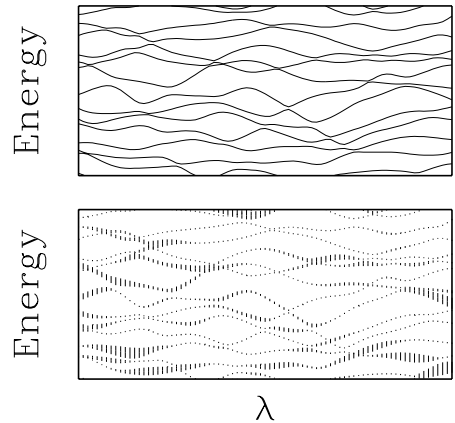, width = 7.4cm}
  \caption{Illustration of ergodic behavior.  The upper square shows how 
  the energy eigenvalues move as a function of $\lambda$.  The lower square 
  is a graphical representation of $S_\alpha(E, \lambda)$.  Each small 
  line segment is centered on an eigenvalue and its lambda value.  The 
  heights are proportional to the overlap intensity with a wavepacket.  The
  level velocities and overlap intensities were produced using a Gaussian
  orthogonal ensemble.}
  \label{level_velocities}
\end{figure}
${\cal C}_\alpha (\lambda)$  thus has a simple form  and the additional advantage
of involving quantities of direct physical interest.  Level velocities
(curvatures also) arise in thermodynamic properties of mesoscopic 
systems~\cite{riedel}, and overlap intensities often arise in the manner 
used to couple into the system~\cite{lane}.  It is the most
sensitive measure of the four possible combinations, the others being the
intensity-curvature, intensity change - curvature and intensity change - level 
velocity correlation coefficients.  The first two are far less sensitive 
measures of eigenstate localization effects, even though curvature 
distributions are affected by localization because of the relative rareness of
being near avoided crossings where curvatures are large.  The last shows no
effect since  intensities will change whether the level is moving up or down. 
These three  measures will not be considered further in this paper, but we did
calculate  them to verify their lack of sensitivity.  

\end{section}

\begin{section}{Ergodicity, Random Waves, and Random Matrix Theory}
\label{rmt}

Semiclassical expressions for wavefunctions have the form
\begin{equation}
  \label{semi}
  \Psi({\bf x}) = \sum_n \ A_n({\bf x}) \exp \left( i S_n({\bf x}) /\hbar - i
  \nu_n \pi/2 \right)
\end{equation}
where $S_n({\bf x})$ is a classical action, $\nu_n$ is a phase index, and
$A_n({\bf x})$ is a slowly varying function given by the square root of a
classical probability.  The classical trajectory underlying each term arrives
at the point ${\bf x}$ with momentum, ${\bf p}_n=\nabla S_n({\bf x})$.  For a
chaotic system, a complete theory leading to an equation of the form of
Eq.~(\ref{semi}) does not exist~\cite{einstein}.  Nevertheless, 
Berry~\cite{berry} conjectured that for the purposes of understanding the
statistical properties of chaotic eigenfunctions, the ergodic hypothesis
implies that the true eigenfunction will appear statistically
equivalent to a large sum of these terms each arriving with a random phase
(since each wave contribution extends over a complicated, chaotic path).  For
systems whose Hamiltonian is a sum of kinetic and potential energies, the
energy surface constraint $\delta(E-H({\bf p},{\bf q}))$ fixes only the
magnitude of the wavevector.  The eigenfunctions therefore appear locally as a
sum of randomly phased plane waves pointing in arbitrary
directions with fixed wavevector $k$.  The central limit theorem asserts such
waves are Gaussian random.  An example is shown in Fig.~(\ref{random}) for a 
two-degree-of-freedom system where the spatial correlations fall off as a Bessel
function, $J_0(kr)$.

If the eigenstates truly possessed these characteristics, then a perturbation 
of the Hamiltonian would have matrix elements that behaved as Gaussian random
variables whose variance depended only on the energy separation of the
two eigenstates, i.e.~an energy-ordered, banded random matrix.  The energy
ordering separates the weakly interacting states, and therefore only the
local structure is of importance here.  The range of the averaging carried out
in the correlation function is taken to be much less than the bandwidth of
such a random matrix.  The ultimate statistical expression of 
this structure is embodied in one of the standard Gaussian ensembles (GE).  We
construct a parametrically varying ensemble $\{\hat H(\lambda)\}$ as
\begin{equation}
  \label{ensemble}
  \hat H (\lambda) = \hat H_0 + \lambda \hat H_1
\end{equation}
where $\hat H_0$ and $\hat H_1$ are independently chosen GE matrices.  Note 
that the sum of two GE matrices is also a GE matrix which thus satisfies our
desire to consider stationary statistical properties as $\lambda$ varies.  

\begin{figure}
  \epsfig{file=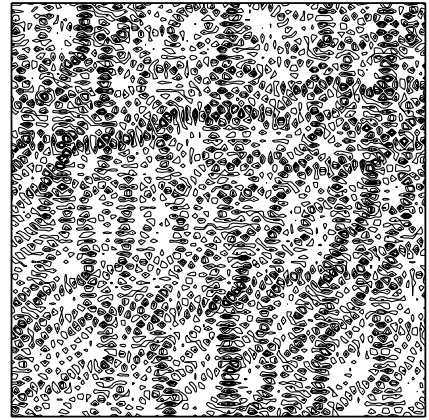, width = 8.2cm}
  \caption{Realization of random wavefunctions in two degrees of freedom.  
  A superposition of 30 plane waves with random direction and phase shift, but 
  fixed magnitude of the wavevector is shown.}
  \label{random}
\end{figure}

It is unnecessary to specify the abstract vector space of $\{\hat
H(\lambda)\}$ (only the dimensionality of the space) in the
definition of the ensemble.  However, $|\alpha\rangle$ has to be overlapped
with the eigenstates, and thus a localized wave packet seemingly must be
specified.  In fact, the specific choice is completely irrelevant because the
GEs are invariant under the set of transformations that diagonalize
them.  $|\alpha\rangle$ can be taken as any fixed vector in the space by
invariance.  The overlaps and level velocities turn out to be independent
over the ensemble since diagonalizing $\{\hat H_0\}$ leaves $\{\hat H_1\}$ invariant
and the level velocities are equal to the diagonal matrix elements of $\hat
H_1$.  With the overbar denoting ensemble averaging,
\begin{equation}
  \label{rmt1}
  \overline{{\cal C}_\alpha(\lambda)} = {\overline{ \left\langle p_{\alpha n}
  {\partial E_n \over \partial \lambda} 
  \right\rangle}_E  \over \sigma_\alpha \sigma_E}
  = {\overline{\left\langle p_{\alpha n} \right\rangle}_E 
  \overline{\left\langle {\partial E_n \over \partial \lambda} \right\rangle}_E 
  \over \sigma_\alpha \sigma_E}=0 
\end{equation}
In fact, it is essential to keep in mind that {\em every} choice of
$|\alpha\rangle$ gives zero correlations within the random matrix
framework.  The existence of even a single $|\alpha\rangle$ in a particular
system that leads to nonzero correlations violates ergodicity.  

It is straightforward to go further and consider the mean square fluctuations 
of
${\cal C}_\alpha (\lambda)$,  
\begin{eqnarray}
  \label{ergodicity}
  \overline{{\cal C}_\alpha (\lambda)^2} 
    &=& {\overline{\left( \left\langle
    p_{\alpha i} {\partial E_i \over \partial \lambda} \right\rangle_E \right)^2} 
    \over \left(\sigma_\alpha \sigma_E\right)^2}
    \nonumber \\ 
  &=& {1 \over \left(N \sigma_\alpha \sigma_E\right)^2} \sum_i^N \sum_j^N \ 
    \overline{p_{\alpha i} {\partial E_i \over \partial \lambda}
    p_{\alpha j} {\partial E_j \over \partial \lambda}}
    \nonumber \\
  &=& {1 \over \left(N \sigma_\alpha \sigma_E\right)^2} \sum_i^N \sum_j^N \ 
    \overline{p_{\alpha i} p_{\alpha j}}
    \overline{{\partial E_i(\lambda) \over \partial \lambda}
    {\partial E_j(\lambda) \over \partial \lambda}}
    \nonumber \\
  &=& {1 \over \left(N \sigma_\alpha \sigma_E\right)^2} \sum_i^N \ 
    \overline{p_{\alpha i}^2}\ \
    \overline{\left({\partial E_i(\lambda) \over \partial \lambda}
    \right)^2}
    = {1\over N}
\end{eqnarray}
where $N$ is the effective number of states used in the energy
averaging.  Again the level velocities are independent of the eigenvector
components.  The $\partial E_j(\lambda) / \partial \lambda =\langle
j|\hat H_1|j\rangle$ and thus the $i\neq j$ terms vanish due to the
independence of the diagonal elements of the perturbation leaving only the
diagonal terms that involve the quantities that respectively enter the variance
of the eigenvector components and the mean square level velocity.  The final
result reflects the  equivalence of ensemble and spectral averaging in the
large-$N$ limit.  Therefore, in ergodically behaving systems, ${\cal C}_\alpha
(\lambda)=0\pm N^{-1/2}$ for every choice of $|\alpha\rangle$. 
Fig.~(\ref{level_velocities}) was made using the orthogonal GE.  It illustrates
a manifestation of ergodicity, i.e.~universal response of the quantum levels
with respect to $\lambda$ and democratic behavior of the overlap intensities.

\end{section}

\begin{section}{Semiclassical Dynamics}
\label{semi_dynamics}

We develop a theory based upon semiclassical dynamics which explains how
nonzero overlap correlation coefficients arise out of the localization properties
of the system.  The theory simply reflects the quantum manifestations of 
finite time correlations in the classical dynamics.  In a chaotic system, 
the classical propagation of $\rho_{\alpha}({\bf p}, {\bf q})$ will relax 
to an ergodic long time average.  However, wave packet revivals in the 
corresponding quantum system earlier than this relaxation time can 
occur~\cite{tl}.  In Heller's original treatment of scars~\cite{heller}, 
he uses arguments based upon these recurrences which occur at finite
times to infer localization in the eigenstates.  

In the correlation function, the intensities, $p_{\alpha n}$, weight most 
heavily the level motions of the group of eigenstates localized near
$\rho_{\alpha}({\bf p}, {\bf q})$, if indeed such eigenstates exist.  
If we construct the Hamiltonian as in Eq.~(\ref{ensemble}) where $\hat H_0$ 
is the unperturbed part, then by first-order perturbation
theory, the level velocities are the diagonal matrix elements of $\hat H_1$
just as in random matrix theory.  We showed in the previous section that in 
random matrix theory these elements weighted with the intensities are zero 
centered.  For a general quantum system the equivalent expectation would be 
fluctuations about the corresponding classical average of the perturbation 
over the microcanonical energy surface, $\delta(E - H({\bf p}, {\bf q}))$.  
In this case, ${\cal C}_\alpha(\lambda) \approx 0$ for all $|\alpha \rangle$.  
On the other hand, the quantum system will fluctuate differently if there is
localization in the eigenstates.  Note that this means some choices of 
$|\alpha \rangle$ will still lead to zero correlations.  It only takes 
one statistically significant nonzero result to demonstrate localization 
conclusively, but to obtain a complete picture, it is necessary to consider 
many $|\alpha \rangle$ covering the full energy surface.

We begin by examining the individual components of the overlap correlation
coefficient, the level velocities and intensities.  Their $\hbar$-dependences 
are derived and also they are shown to be consistent with random matrix theory as
$\hbar \rightarrow 0$.  Finally, the weighted level velocities are discussed.  
We give an estimate based upon a semiclassical theory involving homoclinic 
orbits for the slope of the large intensities.  

\begin{subsection}{Level velocities}

In random matrix theory (RMT) level velocities are Gaussian distributed as 
would also be expected of a highly chaotic system in the small $\hbar$ limit.  
Thus, the mean and variance, $\sigma_E^2$, give a complete statistical 
description in the limiting case and are the most important quantities 
more generally.  Since the purpose of this section is to derive 
their scaling properties, it is better to work with 
dimensionless quantities.  Thus, the dimensionless variance is defined as 
$\tilde{\sigma}_E^2 \equiv \overline{d}^2(E,\lambda)\sigma_E^2$ where
$\overline{d}(E,\lambda)$ is the mean level density which is the 
reciprocal of the mean level spacing.

We begin by following arguments originally employed by Berry and 
Keating~\cite{keating} in which they investigated the level velocities 
normalized by the mean level spacing for classically chaotic systems with 
the topology of a ring threaded by quantum flux.  In order to make the 
discussion self contained we will summarize their basic ideas using their notation 
and then extend their results to include level velocities for any
classically chaotic system.  More recently, Leboeuf and Sieber~\cite{leboeuf}
studied the non-universal scaling of the level velocities using a similar
semiclassical theory.  The $\hbar$-dependence of the average and 
root mean square level velocities for an arbitrary parameter change is derived
and is consistent with the previous works.

The smoothed spectral staircase is
\begin{equation}
  \label{staircase}
  N_\epsilon(E,\lambda) = \sum_n\theta_\epsilon(E - E_n(\lambda))
\end{equation}
and taking the derivative with respect to the parameter, we obtain
\begin{equation}
  \label{deriv_staircase}
  {\partial N_\epsilon(E,\lambda) \over \partial \lambda} 
  = \sum_n\delta_\epsilon(E - E_n(\lambda)){\partial E_n(\lambda) 
  \over \partial \lambda}
\end{equation}
The quantity $\epsilon$ is an energy smoothing term which will be taken
smaller than the mean level spacing.  Our calculations will use
Lorentzian smoothing where
\begin{equation}
  \delta_\epsilon(x) = {\epsilon \over \pi\left(x^2 + \epsilon^2\right)}
\end{equation}
The energy averaging of Eq.~(\ref{deriv_staircase}) yields
\begin{equation}
  \label{staircase_rate}
  \left\langle{\partial N_\epsilon(E,\lambda) 
  \over \partial \lambda}\right\rangle_E = \overline{d}(E,\lambda)
  \left\langle{\partial E_n(\lambda) 
  \over \partial \lambda}\right\rangle_n
\end{equation}
Thus, in order to obtain information about the level velocities, we will 
evaluate the spectral staircase.

The semiclassical construction of the spectral staircase 
is broken into an average part and an oscillating part
\begin{eqnarray}
  \label{semi_staircase}
  N_\epsilon(E,\lambda) &=& \overline{N}(E,\lambda) + \sum_p B_p(E, \lambda)
    \exp\left\{i\left[{S_p(E,\lambda) \over \hbar}\right]\right\} 
    \nonumber \\
  && \times \exp\left\{{-\epsilon T_p(E,\lambda) \over \hbar}\right\}
\end{eqnarray}
The average staircase $\overline{N}(E,\lambda)$ is the Weyl term and to
leading order in $\hbar$ is given by
\begin{equation}
  \label{weyl}
  \overline{N}(E,\lambda) = {1 \over h^d}
  \int\int\theta(E - H({\bf p},{\bf q};\lambda))d{\bf p}d{\bf q}
\end{equation}
This simply states that each energy level occupies a volume $h^d$ in phase
space.  A change in the phase space volume will produce level velocities
due to the rescaling.  We wish to study level velocities created by a
change in the dynamics, not the rescaling.  Hence, without loss of generality
we will require the phase space volume to remain unchanged, so 
$\partial\overline{N}(E,\lambda)/\partial\lambda = 0$. 
The oscillating part of the spectral staircase is a sum
over periodic orbits.  In general, a perturbation will alter the value 
of the classical actions, $S_p$, the periods, $T_p$, and the amplitudes,  
\begin{equation}
  \label{amplitude}
  B_p = {\exp(i\nu_p) \over 2\pi\sqrt{\det(M_p - 1)}}
\end{equation}
where $M_p$ is the stability matrix and $\nu_p$ is the Maslov phase index.
The summation is most sensitive to the changing actions and periods 
because of the associated rapidly oscillating phases, i.e.~the 
division by $\hbar$ in the exponential.  Since the energy smoothing term, 
$\epsilon$, is taken smaller than a mean level spacing, it 
scales at least by $\hbar^d$ and the derivatives of the period vanish as
$\hbar \rightarrow 0$.  Thus, only the derivatives of the
actions are considered, and the oscillating part of the staircase yields
\begin{eqnarray}
  \label{osc_rate}
  \left\langle{\partial N_{osc}(E,\lambda) \over \partial \lambda}\right\rangle_E
    &=& \left\langle\sum_p B_p
    \left[{i \over \hbar}{\partial S_p(E,\lambda) \over \partial \lambda}\right]
    \right. \nonumber \\
  && \times \exp\left\{i\left[{S_p(E,\lambda) \over \hbar}\right]\right\} 
    \nonumber \\
  && \times \left. \exp\left\{{-\epsilon T_p(E,\lambda) \over \hbar}\right\}
    \right\rangle_E
\end{eqnarray}
It has been shown~\cite{ozoriobook} that the change in the action for a periodic 
orbit is
\begin{equation}
  {\partial S_p \over \partial \lambda} 
  = - \int_0^{T_p} {\partial H({\bf p}, {\bf q}; \lambda) \over \partial \lambda} 
  dt
\end{equation}
The above integral is over the
path of the unperturbed orbit and the Hamiltonian can have the form
of Eq.~(\ref{ensemble}) where $H_0$ is the unperturbed part.
Eq.~(\ref{osc_rate}) can be solved without the explicit knowledge of the 
periodic orbits in the $\hbar \rightarrow 0$ limit.  The quantity 
$\partial S_p/\partial \lambda$ is replaced by its average. By the principle
of uniformity~\cite{hannay}, the collection of every periodic orbit covers
all of phase space with a uniform distribution.  Thus, the time integral 
can be replaced by an integral over phase space upon taking the average,
\begin{eqnarray}
  \label{phasespaceintegral}
  \lim_{T \rightarrow \infty} {1 \over T}
    \left\langle {\partial S_p \over \partial \lambda} \right\rangle_p
    &=& {-1 \over V} 
    \int {\partial H({\bf p}, {\bf q}; \lambda) \over \partial \lambda}
    \nonumber \\
  && \times \delta(E - H({\bf p}, {\bf q}; \lambda)) d{\bf p} d{\bf q}
\end{eqnarray}
where $V$ is the phase space volume of the energy surface. 
The above treatment of the average is only valid for the long orbits,
but we may ignore the finite set of short orbits in the sum for
small enough $\hbar$.  $\partial H({\bf p}, {\bf q}; \lambda)/\partial \lambda$ 
is the perturbation of the system that distorts the energy 
surface.  Since the phase space is assumed
to remain constant, then the average change in the actions of the
periodic orbits is zero in the limit of summing over all the orbits.  
If only a finite number of orbits are considered, corresponding to
a finite $\hbar$, then there might
be some residual effect of the oscillating part which will cause
a deviation from RMT.

Continuing to follow  Berry and Keating, the mean square of the counting 
function derivatives can be expressed in terms of the level velocities
\begin{eqnarray}
  \left\langle \left( {\partial N_\epsilon \over \partial \lambda}
    (E, \lambda) \right)^2 \right\rangle_E
    &=& \left\langle \sum_n \sum_m {\partial E_n \over \partial \lambda}
    (\lambda) {\partial E_m \over \partial \lambda} (\lambda) \right. \nonumber \\
  && \times \delta_\epsilon (E - E_n(\lambda)) \nonumber \\ 
  && \left. \times \delta_\epsilon (E - E_m(\lambda)) \right\rangle_E
\end{eqnarray}
For a non-degenerate spectrum, the summation is non-zero only if $n = m$
because of the product of the two delta functions.  Since Lorentzian smoothing 
is applied, then
\begin{equation}
  \label{delta_squared}
  \delta_\epsilon^2(x) \approx {1 \over 2 \pi \epsilon} \delta_{\epsilon/2}(x)
\end{equation}
for $\epsilon \ll \overline{d}^{-1}$.  Thus we have 
\begin{equation}
  \left\langle \left( {\partial N_\epsilon \over \partial \lambda}
  (E, \lambda) \right)^2 \right\rangle_E
  = {\overline{d} \over 2 \pi \epsilon} \left\langle \left( 
  {\partial E_n \over \partial \lambda} (\lambda) \right)^2 \right\rangle_n
\end{equation}
The final result will be independent of $\epsilon$ and the type of
smoothing, i.e.~Lorentzian or Gaussian.  Using the $\lambda$ derivative
of Eq.~(\ref{semi_staircase}), the dimensionless level velocities are
\begin{eqnarray}
  \label{berry_auto}
  \tilde{\sigma}_E^2 &=& {2 \pi \epsilon \overline{d} \over \hbar^2} 
    \left\langle\sum_p\sum_{p^\prime} |B_p B_{p^\prime}| 
    {\partial S_p \over \partial \lambda}
    {\partial S_{p^\prime} \over \partial \lambda} \right. \nonumber \\
  &\times& \left. \exp\left\{i\left[{S_p - S_{p^\prime} \over \hbar}\right]\right\}
    \exp\left\{{-\epsilon \over \hbar}\left[T_p + T_{p^\prime}\right]\right\}
    \right\rangle_E
\end{eqnarray}

The diagonal and off-diagonal contributions are separated, so
\begin{equation}
  \label{berry_auto2}
  \tilde{\sigma}_E^2 = \tilde{\sigma}_{E,diag}^2 + \tilde{\sigma}_{E,off}^2
\end{equation}
As $\hbar \rightarrow 0$, the phase of the exponential oscillates rapidly and 
averages out to be zero unless $S_p = S_{p^\prime}$.  We will assume that this 
occurs rarely except when $p = p^\prime$.  The product 
$(\partial S_p / \partial \lambda)(\partial S_{p^\prime} / \partial \lambda)$ 
can take on both positive and negative values.  This also helps to reduce the 
contributions of the off-diagonal terms.  For a more complete discussion of the 
diagonal vs.~off-diagonal terms see~\cite{olm}.  We will only present the results 
for the diagonal terms, since the correlations between the actions of different 
orbits is not known but should not alter the leading $\hbar$-dependence.

The diagonal contribution is
\begin{equation}
  \label{diag}
  \tilde{\sigma}_{E,diag}^2 = {2 \pi \epsilon \overline{d} g \over \hbar^2}
  \left\langle\sum_p |B_p|^2\left({\partial S_p \over \partial \lambda}\right)^2
  \exp\left\{{-2\epsilon T_p \over \hbar}\right\}
  \right\rangle_E
\end{equation}
The factor $g$ depends on the symmetries of the system.  For systems with
time-reversal invariance $g = 2$ and without time-reversal symmetry $g = 1$.
The precise values of $\partial S_p/\partial \lambda$ are specific to
each periodic orbit rendering the sum difficult to evaluate precisely. A 
statistical approach is possible though which generates a relationship 
between the sum and certain correlation decays.  Hence, the quantity 
$(\partial S_p/\partial \lambda)^2$ in Eq.~(\ref{diag}) is replaced by its average,
\begin{eqnarray}
  \label{rms_action}
  \left\langle \left( {\partial S_p \over \partial \lambda} \right)^2 
    \right\rangle_p  &=& \int_0^T \int_0^T \left\langle
    {\partial H({\bf p}(t), {\bf q}(t); \lambda) \over \partial \lambda} 
    \right. \nonumber \\
  &\times& \left. {\partial H({\bf p}(t^\prime), {\bf q}(t^\prime); \lambda) 
    \over \partial \lambda}
    \right\rangle_p dt^\prime dt \nonumber \\
  &=& 2 \int_0^T \int_t^T \left\langle
    {\partial H({\bf p}(t), {\bf q}(t); \lambda) \over \partial \lambda}
    \right. \nonumber \\
  &\times& \left. {\partial H({\bf p}(t^\prime + t), {\bf q}(t^\prime + t); \lambda) 
    \over \partial \lambda}
    \right\rangle_p dt^\prime dt
\end{eqnarray}
Long orbits increasingly explore the available phase space on an ever finer 
scale.  As the time between two points in a chaotic system goes to infinity, 
then they become uncorrelated from each other.  This is a consequence 
of the mixing property,
\begin{equation}
  \label{mixing}
  \left\langle f(0)f(t) \right\rangle_p \rightarrow 0
\end{equation}
This property is independent of the placement of the two points, i.e.~the
two points can lie on the same orbit as long as the time between the
points increases to infinity. 
Thus, by the central limit theorem, $\partial S_p/\partial \lambda$ will
be Gaussian distributed for the sufficiently long periodic orbits.  
The time dependence of Eq.~(\ref{rms_action})
is approximated by a method discussed by Bohigas {\it et~al.}~\cite{bgos}.
They define
\begin{equation}
  \label{k(e)}
  K(E) = \int_0^\infty \left\langle {\partial H({\bf p}(0), {\bf q}(0); \lambda) 
  \over \partial \lambda}
  {\partial H({\bf p}(t), {\bf q}(t); \lambda) 
  \over \partial \lambda} \right\rangle_p dt
\end{equation}
which can be evaluated in terms of properties of the perturbation.
The variance of the actions in the limit of long periods becomes
\begin{equation}
  \label{k(e)2}
  \left\langle \left( {\partial S_p \over \partial \lambda} \right)^2 
  \right\rangle_p \approx 2K(E)T
\end{equation}  
Applying the Hannay and Ozorio de Almeida sum rule~\cite{hannay}, the following 
substitution is made
\begin{equation}
  \label{sumrule}
  \sum_p|B_p|^2 \cdots \rightarrow {1 \over 2\pi^2}\int_0^\infty {dT \over T} 
  \cdots
\end{equation}
Hence, the diagonal contribution is
\begin{eqnarray}
  \label{diag2}
  \tilde{\sigma}_{E,diag}^2 &\approx& {\epsilon \overline{d} g \over \pi\hbar^2}
    \int_0^\infty {1 \over T}\left( 2K(E)T \right)
    \exp\left\{{-2\epsilon T \over \hbar}\right\} dT \nonumber \\
  &\approx& {gK(E)\overline{d} \over \pi \hbar} \nonumber \\
  &\propto& \hbar^{-(d+1)}
\end{eqnarray}
The variance of the level velocities on the scale of a mean spacing grows 
$\hbar^{-1}$ faster than the density of states as the semiclassical limit
$(\hbar \rightarrow 0)$ is approached; see numerical tests performed on the 
stadium in the next section.  

The exact level velocities are perturbation dependent and cannot be determined 
without specific knowledge of the system (i.e.~the evaluation of $K(E)$).
$K(E)$ is a classical quantity that contains dynamical information about the
periodic orbits.  It should scale as the reciprocal of the Lyapunov 
exponent~\cite{goldberg}.  Leboeuf and Sieber derived $K(E)$ for billiards
where the perturbation is a moving boundary.  In this case $K(E)$ depends upon
the autocorrelation function and the fluctuations of the number of bounces.
For maps $K(E)$ is an action velocity diffusion coefficient~\cite{lakshminarayan}.
$\{\partial S_p / \partial \lambda\}$ being Gaussian distributed is linked to the
level velocities being Gaussian distributed as in RMT.  If the 
$\{\partial S_p / \partial \lambda\}$ are not Gaussian distributed by the 
Heisenberg time, then one should not expect the level velocities to be consistent 
with RMT; again see the stadium results ahead.

\end{subsection}

\begin{subsection}{Overlap intensities}
\label{overlapintensities}

Now we investigate the overlap intensities and derive a semiclassical
expression for the $\hbar$-scaling of the root mean square.  Eckhardt
{\it et al.}~\cite{eckhardt} developed a semiclassical theory based
on periodic orbits to obtain the matrix elements of a sufficiently 
smooth operator.  However, the projection operator of interest here, 
$|\alpha\rangle\langle\alpha|$, is not smooth on the scale of $\hbar$ 
for Gaussian wave packets.  Thus, their stationary phase approximations
do not apply, in principle, to the oscillating part of the strength function.
In Berry's work on scars~\cite{berryscar}, he used Gaussian smoothing of
the Wigner transform of the eigenstates to obtain a semiclassical expression
for the strength of the scars.  His approach led to a sum over periodic
orbits.  We will use the energy Green's function similar to Tomsovic and
Heller in~\cite{th1} where they derived the autocorrelation function using
the time Green's function and gave results for the strength functions as well.  
This technique results in a connection between the overlap intensities and the 
return dynamics, namely the homoclinic orbits.

For completeness, we present the smooth part of the strength function
which is easily obtained from the zero-length trajectories,   
\begin{equation}
  \label{diag_matrix}
  \overline{S}_{\alpha}(E, \lambda) = {1 \over h^d} \int A({\bf q},{\bf p}) 
  \delta(E - H({\bf q},{\bf p}))d{\bf q}d{\bf p}
\end{equation}
$A({\bf q},{\bf p})$ is the Wigner transform of the Gaussian
wave packet and is given by
\begin{equation}
  \label{wigner}
  A({\bf q},{\bf p}) = 2^d \exp \{ -({\bf p} - {\bf p}_\alpha)^2 \sigma^2 
  / \hbar^2 - ({\bf q} - {\bf q}_\alpha)^2 / \sigma^2 \}
\end{equation}
The above results were previously used by Heller~\cite{leshouches} 
in the derivation the envelope of the strength function and does not contain
any information about the dynamics of the system.

The oscillating part of the strength function, on the other hand, includes
dynamical information,
\begin{equation}
  \label{strength_osc}
  S_{\alpha,osc}(E, \lambda) = {-1 \over \pi} {\mathrm Im} 
  \int \langle \alpha | {\bf q} \rangle 
  G({\bf q}, {\bf q}^\prime;E)
  \langle {\bf q}^\prime | \alpha \rangle d{\bf q} d{\bf q}^\prime
\end{equation}
where
\begin{eqnarray}
  \label{green}
  G({\bf q}, {\bf q}^\prime;E)
    &=& {1 \over i\hbar (2\pi i \hbar)^{(d - 1)/2}} \sum_{j} |D_s|^{1/2} \nonumber \\ 
  && \times e^{ i [S_j({\bf q}, {\bf q}^\prime;E)/\hbar - \nu_j^\prime\pi/2]}
\end{eqnarray}
is the semiclassical energy Green's function.
The above sum is over all paths that connect ${\bf q}$
to ${\bf q}^\prime$ on a given energy surface $E$.
The action is quadratically expanded about each reference trajectory;
see Appendix~\ref{A} for details.  The initial and final points 
(${\bf q}_i$ and ${\bf q}_f$) of the reference trajectories are 
obtained by considering the evolution of the wave packet.  Nearby points 
will behave similarly for short times.  Thus, the phase space can be 
partitioned into connecting areas.  As the time is increased the number 
of partitions grow and the size of their area shrinks.  The reference 
trajectories are the paths that connect the partitions.  The autocorrelation
function in~\cite{th1} has the same form as Appendix~\ref{A} where the
paths that contribute to the saddle points are the orbits homoclinic to
the centroid of the Gaussian wave packet so that ${\bf q}_i$ and ${\bf q}_f$ 
lie on the intersections of the stable and unstable manifolds. The result from 
Appendix~\ref{A} is
\begin{eqnarray}
  \label{semi_strength}
  S_{\alpha,osc}(E) &=& {\sigma \over \pi^{1/2} \hbar} {\mathrm Re} \sum_j
    \left( {\det \tilde{{\bf A}}^{21} \over \det{\bf A}} \right)^{1/2} \\
  &\times& \left( { 1 \over |\dot{q}^{(N)}| |\dot{q}^{\prime (N)}|} \right)^{1/2}
    f_j({\bf q}_f, {\bf q}_i) e^{i S_j({\bf q}_f, {\bf q}_i; E) / \hbar}
    \nonumber 
\end{eqnarray}
where
\begin{eqnarray}
  f_j({\bf q}_f, {\bf q}_i) 
    &=& \exp \left\{{1 \over 4}{\bf b} \cdot {\bf A}^{-1} \cdot {\bf b} 
    - {i \over \hbar} {\bf p}_\alpha \cdot ({\bf q}_f - {\bf q}_i) 
    \right. \\
  && \left. \times - {({\bf q}_f - {\bf q}_\alpha)^2 \over 2 \sigma^2}
    - {({\bf q}_i - {\bf q}_\alpha)^2 \over 2 \sigma^2}
    - {i \nu^\prime_j \pi \over 2} \right\} \nonumber
\end{eqnarray} 
The function $f_j({\bf q}_f, {\bf q}_i)$ in the above equation is a damping 
term which depends on the end points of the homoclinic orbits. 
Only orbits which approach the center of the Gaussian wave packet 
in phase space will contribute to the sum.  The time derivatives of the
parallel coordinates are evaluated at the saddle points which are near
the centroid of the Gaussian, so we may set 
$|\dot{q}^{(N)}| \approx |\dot{q}^{\prime (N)}| \approx |p_\alpha|/m$.
The sum  over homoclinic orbits used for the autocorrelation function 
in~\cite{th1} converged well to the discrete quantum strength function 
when only those orbits whose period did not exceed the Heisenberg time, 
$(\tau_{H} = 2 \pi \hbar \overline{d}(E,\lambda))$, were included.  
As happened with the periodic orbits and the level velocities, in order 
to evaluate Eq.~(\ref{semi_strength}) the homoclinic orbits and their 
stabilities must be computed rendering the sum tedious to evaluate
precisely as done in~\cite{th1}. 

By taking a statistical approach we can gain some insight into the 
workings of this summation.  The variance of the intensities are obtained 
by a similar fashion as the level velocities. Using Eq.~(\ref{strengthfunction}) 
and Eq.~(\ref{delta_squared}), we have
\begin{equation}
  \left \langle S_{\alpha,osc}^2(E,\lambda) \right \rangle_E 
  = {\overline{d} \over 2 \pi \epsilon} \sigma_\alpha^2
\end{equation}
Since the square of the strength function is a product of two delta
functions, an energy smoothing term is required.  After making the 
diagonal approximation, we obtain
\begin{eqnarray}
  \label{sigma_alpha^2}
  \sigma_{\alpha,diag}^2 &=& {2 \pi \epsilon g\over \overline{d}} \left\langle
    \sum_j {m^2 \sigma^2 \over \pi \hbar^2}
    \left| {\det \tilde{{\bf A}}^{21} \over \det{\bf A}} \right| 
    \right. \nonumber \\
  && \left. \times {|f_j({\bf q}_f, {\bf q}_i)|^2 \over |p_\alpha|^2} 
    e^{-2\epsilon T_j/\hbar} \right \rangle_E
\end{eqnarray}
A classical sum rule is applied to the above sum for special cases including
two-dimensional systems; see Appendix~\ref{B} for the details.  Thus,
\begin{eqnarray}
  \label{var_strength}
  \sigma_{\alpha,diag}^2
    &\approx& {2 \epsilon g m^2 \sigma^2 \over \hbar^2 \overline{d} |p_\alpha|^2} 
    \int \exp\{-2\epsilon T/\hbar \} dT \nonumber \\
  &\approx& {g m^2 \sigma^2 \over \hbar \overline{d} |p_\alpha|^2}
\end{eqnarray}
Setting $\sigma \propto \hbar^{1/2}$ which shrinks the momentum and position
uncertainties similarly, the $\hbar$-scaling of $\sigma_{\alpha,diag}^2$ is 
$\hbar^d$; see numerical tests of the stadium in the next section.

Assuming that the amplitudes of the wavefunctions are Gaussian random, then 
the RMT result for strength functions is a Porter-Thomas distribution
which has a variance that is proportional to the square of its average.
The average strength function, Eq.~(\ref{diag_matrix}), scales as 
$(\sigma/\hbar)^d$ for Hamiltonians which can be locally expanded 
as a quadratic.  Therefore, with $\sigma \propto \hbar^{1/2}$
the variance of the strength function, Eq.~(\ref{var_strength}), scales as 
the square of the average and is consistent with RMT.

\end{subsection}

\begin{subsection}{Weighted level velocities}

A semiclassical treatment of the overlap correlation coefficient defined in 
Eq.~(\ref{correlator}) is now developed. As stated in the introduction, the
companion paper~\cite{lakshminarayan2} presents the semiclassical theory for 
maps.  We stress that in the preceding subsections and in what follows is for 
conservative Hamiltonian systems.  Here, the $\hbar$-dependence of the average 
overlap correlation coefficient is established and a semiclassical argument for
the existence of nonzero correlations is presented.

\begin{subsubsection}{Actions of homoclinic orbits}
\label{homo_actions}

To calculate the overlap correlation coefficient, the rate of change of the
actions for homoclinic orbits will be necessary.  As discussed earlier,
this was accomplished for periodic orbits~\cite{ozoriobook}.  
We extend these results to include 
the actions of homoclinic orbits.  Homoclinic orbits have infinite periods
causing their actions to become infinite.  We are 
interested in the limiting difference of the action, ${\mathcal S}^{(p)}_j$, 
between the $j^{th}$ homoclinic orbit and repetitions of its corresponding 
periodic orbit $p$.  The difference is finite and is equal to the area bounded 
by the stable and unstable manifolds with intersection
at the $j^{th}$ homoclinic point in a Poincar\'{e} map.  
${\mathcal S}^{(p)}_j$ provides information about the additional phase 
gathered by the homoclinic orbit. The action of the $j^{th}$ homoclinic orbit as 
$n \rightarrow \infty$ in the time interval $(-nT_p, nT_p)$ is
\begin{equation}
  \label{homo_action}
  S^{(p)}_{n,j} \rightarrow 2nS_p + {\mathcal S}^{(p)}_j
\end{equation}
where $T_p$ and $S_p$ are the period and action of the periodic orbit,
respectively. As a consequence of the Birkhoff-Moser theorem~\cite{moser}, if the 
Poincar\'{e} map is invertible and analytic, then there exist infinite families 
of periodic orbits that accumulate on a homoclinic orbit.  It is thus
possible to estimate the action of the homoclinic orbit by these periodic
orbits whose action is given by~\cite{ozorio}
\begin{equation}
  \label{action_difference}
  \alpha^{(p)}_{n,j} = nS_p + {\mathcal S}^{(p)}_j - s^{(p)}_{n,j}
\end{equation}
where $s^{(p)}_{n,j}$ is the difference in action between a path defined by 
${\mathcal S}^{(p)}_j$ along the stable and unstable manifolds and the 
path of the new periodic orbit in a Poincar\'{e} map.  $s^{(p)}_{n,j}$ depends 
exponentially on $n$, so as $n \rightarrow \infty$, $\alpha^{(p)}_{2n,j}$ 
approaches the action of the homoclinic orbit.  Thus, in the limit of 
large $n$, ${\mathcal S}^{(p)}_j$ is approximated by the difference between 
two periodic orbits 
(i.e.~${\mathcal S}^{(p)}_j \approx \alpha^{(p)}_{n,j} - nS_p$). 
Hence, the change in ${\mathcal S}^{(p)}_j$ due to a small perturbation is 
calculated as in~\cite{ozoriobook},
\begin{eqnarray}
  \label{action_pert}
  \Delta {\mathcal S}^{(p)}_j &=& -\Delta\lambda\int_{\alpha^{(p)}_{n,j}}
    {\partial H({\bf p}, {\bf q}; \lambda) \over \partial \lambda}dt 
    \nonumber \\
  && + n\Delta\lambda\int_{S_p}
    {\partial H({\bf p}, {\bf q}; \lambda) \over \partial \lambda}dt 
    + O(\lambda^2)
\end{eqnarray}
where the integrals are over the unperturbed periodic orbits.  The differences,
$s^{(p)}_{n,j}$, can be made smaller than the second order term in 
Eq.~(\ref{action_pert}) by taking $n$ large enough.  Interchanging the order
of integration and differentiation, the integrals reduce to the unperturbed
energy times the derivative of the orbit period with respect to the parameter,
\begin{equation}
  \label{homo_action_times}
  \Delta {\mathcal S}^{(p)}_j \approx 
  -\Delta\lambda E_p{\partial \over \partial \lambda}
  \left(T_{\alpha^{(p)}_{n,j}} - nT_p\right)
\end{equation}
The orbit period $T$ can be expressed as $\partial S/\partial E$.
Thus, the difference of the two periods as $n \rightarrow \infty$ is
\begin{equation}
  \label{homo_action_diff_times}
  T_{\alpha^{(p)}_{n,j}} - nT_p = {\partial \alpha^{(p)}_{n,j} \over \partial E}
  - n {\partial S_p \over \partial E} 
  \approx {\partial {\mathcal S}^{(p)}_j \over \partial E}
\end{equation}
Hence,
\begin{eqnarray}
  \label{homo_action_pert}
  \Delta {\mathcal S}^{(p)}_j 
    &\approx& -\Delta\lambda{\partial \over \partial \lambda}
    \left(E_p{\partial {\mathcal S}^{(p)}_j \over \partial E}\right) \nonumber \\  
  &\approx& -\Delta\lambda\int_{{\mathcal S}^{(p)}_j}
    {\partial H({\bf p}, {\bf q}; \lambda) \over \partial \lambda}dt
\end{eqnarray}
Note that the integral is over the unperturbed path along the stable
and unstable manifolds.  For zero correlations,
$\Delta {\mathcal S}^{(p)}_j$ must be ``randomly'' distributed about zero.

If enough time is allowed, then for ergodic systems the set of 
all homoclinic orbits for a given energy will come arbitrarily
close to any point in phase space on that energy surface.  Thus, an integral
over phase space on the original energy surface can be substituted for the 
time integral,
\begin{eqnarray}
  \label{aver_action}
  \lim_{T \rightarrow \infty} {1 \over T} 
    \left\langle\Delta {\mathcal S}^{(p)}_j\right\rangle_j 
    &=& {-\Delta \lambda \over V}
    \int \delta\left(E - H\left({\bf p}, {\bf q}; 0\right)
    \right) \nonumber \\
  && \times {\partial H\left({\bf p}, {\bf q}; \lambda\right) \over \partial \lambda}
    d{\bf p} d{\bf q}
\end{eqnarray}
Since the density of states are kept constant, the perturbations
fluctuate about zero and the integral vanishes.  Hence, the average
change in the actions will be zero.  The mean square fluctuations of the actions,
\begin{eqnarray}
  \label{mean_sq_action}
  \left\langle \left( \Delta {\mathcal S}^{(p)}_j \right)^2\right\rangle_j 
    &=& \left(\Delta \lambda\right)^2 
    \int_0^{\tau_H} \int_0^{\tau_H} \left\langle 
    {\partial H\left({\bf p}, {\bf q}; \lambda\right) \over \partial \lambda}
    \right. \nonumber \\
  && \left. \times {\partial H\left({\bf p}^\prime, {\bf q}^\prime; \lambda\right) 
    \over \partial \lambda} \right\rangle_j dt dt^\prime
\end{eqnarray}
approaches a Gaussian distribution by the central limit theorem via the
same reasoning as that for periodic orbits.  Again we can define
$K_{hom}(E)$ as in Eq.~(\ref{k(e)}), except now the average is over
homoclinic orbits and instead of integrating to infinity we only integrate
to the Heisenberg time to be consistent with the range of the sum in 
Eq.~(\ref{semi_strength}).  It is the short time dynamics that dominate.  
Long time correlations will average to zero by the mixing property 
(Eq.~(\ref{mixing})).  Thus, the variance of the actions becomes
\begin{equation}
  \left\langle\ \left( {\partial {\mathcal S}^{(p)}_j \over \partial \lambda} 
  \right)^2 \right\rangle_j \approx 2 K_{hom}(E) T
\end{equation}
$K_{hom}(E)$ will approach $K(E)$ in the semiclassical limit 
($\tau_H \rightarrow \infty$).

\end{subsubsection}

\begin{subsubsection}{Overlap intensity-level velocity correlation coefficient}

In the previous two subsections, we have examined the pieces that 
constitute the overlap correlation coefficient.  The semiclassical 
theories of the level velocities and the intensities are now combined
to construct a semiclassical theory for the weighted level velocities.
The numerator of the overlap correlation coefficient is proportional to 
the energy averaged product of the intensities and level velocities,
\begin{eqnarray}
  \left\langle S_\alpha(E, \lambda) 
    {\partial N(E,\lambda) \over \partial \lambda} \right\rangle_E
    &=& \left\langle \sum_n \sum_m p_{\alpha n}(\lambda) 
    {\partial E_m \over \partial \lambda} \right. \nonumber \\ 
  && \left. \times \delta_\epsilon(E - E_n) \delta_\epsilon(E - E_m)
    \right\rangle_E \nonumber \\
  &=& {\overline{d} \over 2 \pi \epsilon} \left\langle
    p_{\alpha n}(\lambda)
    {\partial E_n \over \partial \lambda} 
    \right\rangle_n \nonumber \\
  &=& {\overline{d} \over 2 \pi \epsilon} \tilde{\cal C}_\alpha(\lambda)
\end{eqnarray}
Lorentzian smoothing was again employed, Eq.~(\ref{delta_squared}), and we've 
defined $\tilde{\cal C}_\alpha(\lambda)$ to be the numerator of the overlap 
correlation coefficient (without the division of the rms level velocities 
and intensities).  By the definition of the overlap correlation coefficient 
only the oscillating part of the level velocities and the intensities are 
considered.  Using the derivative with respect to lambda of 
Eq.~(\ref{semi_staircase}) and Eq.~(\ref{semi_strength}) the numerator becomes
\begin{eqnarray}
  \tilde{\cal C}_\alpha(\lambda) &=& {2 \pi \epsilon \over \overline{d}}
     \left \langle {\mathrm Re} \sum_j \sum_p {m \sigma B_p \over \pi \hbar^2}
     \left( {\det \tilde{{\bf A}}^{21} \over \det {\bf A}} \right)^{1/2} \right. \\
   && \left. \times {f_j({\bf q}_f, {\bf q}_i) \over |p_\alpha|}
     \left( {\partial S_p \over \partial \lambda} \right)
     e^{i(S_j - S_p) / \hbar - \epsilon (T_j + T_p) / \hbar}
     \right \rangle_E \nonumber
\end{eqnarray}
Because of the rapidly oscillating phases, the energy averaging will result 
in zero unless $S_j \approx S_p$.  As stated earlier, for every homoclinic 
orbit there is a periodic orbit that comes infinitesimally 
close to it.  The same periodic orbit's action can be nearly equal to the 
actions of different segments of the same homoclinic orbit.
Thus, a diagonal approximation is used for the homoclinic
segments
\begin{eqnarray}
  \tilde{\cal C}_\alpha(\lambda)
    &\approx& {2 \pi \epsilon g \over \overline{d}} \left \langle {\mathrm Re} 
    \sum_j {m \sigma B_j \over \pi \hbar^2}
    \left( {\det \tilde{{\bf A}}^{21} \over \det {\bf A}} \right)^{1/2}
    e^{-2\epsilon T_j / \hbar} \right. \nonumber \\
  && \left. \times \left( {\partial S_j \over \partial \lambda} \right) 
    {f_j({\bf q}_f, {\bf q}_i) \over |p_\alpha|} \right \rangle_E
\end{eqnarray}
Upon applying the sum rule for two-dimensional systems and other special cases,
Eq.~(\ref{homosumrule}), we have
\begin{eqnarray}
  \label{correlation1}
  \tilde{\cal C}_\alpha(\lambda)
    &\approx& {\epsilon g m \sigma \over \pi \overline{d} |p_\alpha|} {1 \over \hbar^2} 
    \int_{0}^{\infty} e^{-2 \epsilon T/\hbar} \nonumber \\
  && \times \left \langle \left( {\partial S_j \over \partial \lambda} \right)
    f_j({\bf q}_f, {\bf q}_i) \right \rangle_j dT
\end{eqnarray}
The changes in action of the homoclinic excursions are now weighted by the 
$f_j({\bf q}_f, {\bf q}_i)$'s.  Without the additional weighting the average 
in the changes in the action would be zero for all positions of the Gaussian
wave packet.  

In~\cite{tomsovic}, a heuristic argument for the direction of the weighted level
velocities was given.  The argument basically states that the energy surface
changes with the parameter such that the action changes are minimized.  
Eq.~(\ref{correlation1}) differs from~\cite{tomsovic} in that the proper 
weightings, $f_j({\bf q}_f, {\bf q}_i)$, of the homoclinic orbits are derived here,
and the action changes are not correlated with the inverse periods.
Also, in~\cite{tomsovic} the homoclinic orbits were strictly cut-off
at the Heisenberg time whereas here there is an exponential decay on the
order of the Heisenberg time with the energy smoothing term, $\epsilon$,
equal to $\hbar/\tau_H$~\cite{keating}.  One reason that~\cite{tomsovic}
reported such good results is that since the number of homoclinic segments 
proliferate exponentially, most of the included segments occurred near the
Heisenberg time and the expression in Eq.~(\ref{correlation1}) is divided by
the Heisenberg time (i.~e.~multiplied by $\epsilon$).  

As $\hbar \rightarrow 0$ $(\tau_H \rightarrow \infty)$, the integral in 
Eq.~(\ref{correlation1}) would be dominated by the subset of 
$\{\partial S_j / \partial \lambda\}$ associated with very long orbits
and would decouple from the weightings.  For small enough $\hbar$, as 
previously stated, the $\partial S_j / \partial \lambda$ for these orbits 
would approach a zero-centered Gaussian density, and the integral would vanish.  
In other words, we could use arguments analogous to those underlying
Eq.~(\ref{phasespaceintegral}) to write
\begin{eqnarray}
  \tilde{\cal C}_\alpha(\lambda)
    &\approx& {-g m \sigma \over \pi \overline{d} |p_\alpha|}
    \int^\infty_0 dT \, e^{-2T/\tau_H} {T \over \tau_H} 
    {F \over V} \nonumber \\
  && \times \int d{\bf p} d{\bf q}
    {\partial H({\bf p},{\bf q};\lambda) \over \partial \lambda} 
    \delta(E - H({\bf p},{\bf q};\lambda)) 
\end{eqnarray}
where $F$ is the phase space average of the weightings. Note that the phase 
space average of $\partial H({\bf p},{\bf q};\lambda)/\partial \lambda$ vanishes 
excluding an irrelevant drift of levels, so the RMT prediction of 
$\tilde{\cal C}_\alpha(\lambda)$ is recovered for $\hbar$ small enough.

The leading order in $\hbar$ correction to this is more difficult to ascertain.  
$\hbar$ enters into the exponential in the integral for the energy smoothing, but
not for the classical decay of the action changes.  Upon taking the integral,
this yields two competing terms for the $\hbar$-scaling which may depend
upon the region of phase space the correlation is taken in.  The numerics also
show a large fluctuation of the scaling in the stadium (see the next section).

\end{subsubsection}

\end{subsection}

\end{section}

\begin{section}{Stadium Billiard}
\label{stadium_billiard}

\begin{figure}
  \begin{center}
  \epsfig{file=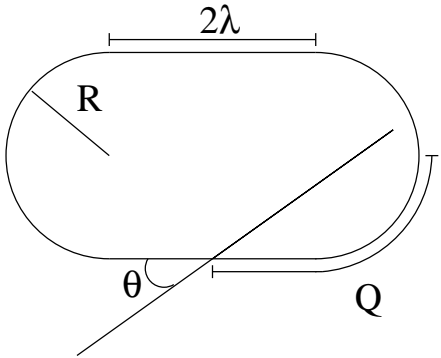, height=5.8cm, angle=270}
  \end{center}
  \caption{Birkhoff coordinates for the stadium billiard.  The position coordinate
  can be taken to start anywhere along the perimeter.  Here we have chosen the
  origin to be the middle of the right semicircle.}
  \label{Birkhoff}
\end{figure}

In this section the semiclassical theories just presented and the numerical 
results from the stadium billiard are compared. The stadium billiard, which 
was proven by Bunimovich~\cite{bunimovich} to be classically chaotic, has
become a paradigm for studies of quantum chaos.  It is defined as a 
two-dimensional infinite well with the shape pictured in Fig.~(\ref{Birkhoff}).  
We continuously vary the side length, $2 \lambda$, while altering the radii 
of the endcaps, $R$, to keep the area of the stadium a constant.  Throughout
this section, the level velocities and intensities are evaluated for
a stadium with $\lambda = R = 1$.  For billiards the average number 
of states below a given energy, $E$, is approximately 
$\overline{N}(E) \approx mAE/2 \pi \hbar^2$ where $A$ is the area
of the billiard.  This is the first term in an asymptotic series
in powers of $\hbar$.  The density of the states, $d\overline{N}/dE$, is then
a constant not depending on $\lambda$ to the lowest power of $\hbar$ if the 
area remains the same.

\begin{figure}
  \epsfig{file=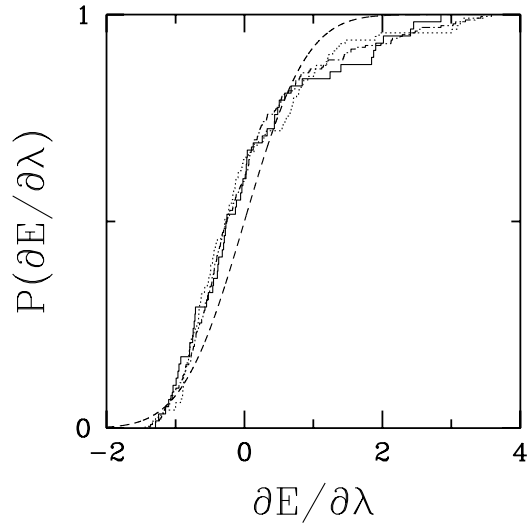, width=8.2cm}
  \caption{Distribution of the level velocities for a stadium billiard.
  The solid line is the lowest energy range, the dotted line is the middle
  energy range and the dash-dot line is the highest energy range.  The RMT
  result is given by the dashed line.  The level velocities have been rescaled to
  zero mean and unit variance.}
  \label{distribution}
\end{figure}

We will examine three different energy regimes for the stadium.  Since
billiards are scaling systems, this will correspond to three different
values of $\hbar$.  The energy regimes are separated by a factor of
four in energy or conversely a factor of one half in $\hbar$.  Twice
as many states are taken in each successive energy regime so that the 
averages will incorporate the same relative size interval in energy as $\hbar$
is decreased.  This corresponds to the increase in the density of states for
varying $\hbar$.

The distributions of the level velocities for all three energy regimes
are shown in Fig.~(\ref{distribution}) along with the random matrix theory
prediction.  The skewness occurs because of a class of marginally
stable orbits in the stadium.  These orbits are the bouncing ball 
orbits which only strike the straight edges.  Their contribution do not 
seem to decrease as the semiclassical limit is approached though they should
once $\hbar$ is sufficiently small.  There is no
clear trend for the level velocity distribution to approach Gaussian
behavior.  The root mean square of the level
velocities also deviates from our calculations of the $\hbar$-scaling
in Section~\ref{semi_dynamics} (Fig.~(\ref{velocity_variance})).  This
is again explained by the bouncing ball orbits whose effects are missing 
from the trace formula.  Quantizing only
these orbits using WKB yields a dimensionless level velocity scaling of 
$\hbar^{-2}$, while the trace formula gives a scaling of $\hbar^{-3/2}$.  
The numerical results give a scaling of approximately $\hbar^{-1.8}$ 
which lies in between the two suggesting that the marginally stable 
orbits significantly effect the level velocities.

\begin{figure}
  \epsfig{file=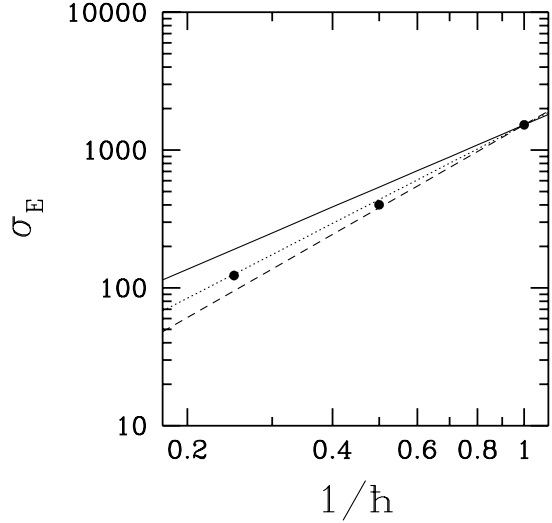, width=8.2cm}
  \caption{Root mean square of the level velocities as a function of $1/\hbar$.
  The solid line is the theoretical value from Section~\ref{semi_dynamics}
  and the dashed line is the WKB results for the bouncing ball motion.
  The best fit line through the stadium results is the dotted line.}
  \label{velocity_variance}
\end{figure}

To study the intensities, the eigenstates must first be constructed.
Bogomolny's transfer operator method~\cite{bogomolny_top} 
was used to find the eigenstates.  This method uses
a $(d - 1)$ dimensional surface of section.  A convenient choice 
is the boundary of the stadium (Fig.~(\ref{Birkhoff})).  The generation 
of a full phase space picture of the stadium would otherwise require four 
dimensions, two positions and two momenta.  The position coordinate is 
measured along the perimeter and the momentum coordinate is defined by 
$\cos \theta$.  The classical dynamics have a quantum analog that uses source 
points on the boundary.  Thus, all of the eigenfunction's localization behavior 
can be explored using wave packets defined in these coordinates.
A coherent state on the boundary is a one-dimensional Gaussian wave
packet; see the lower figure in Fig.~(\ref{gaussian}).  The corresponding 
wave packet in the interior of the stadium can be generated by a 
Green's function and is shown in the upper figure of Fig.~(\ref{gaussian}).
For billiards the Green's function is proportional to a zero$^{th}$
order Hankel function of the first kind, $H_0^{(1)}(kr)/2i\hbar^2$.
The centroid of the Gaussian wave packet is moved along the boundary 
and its momentum is changed according to the Birkhoff coordinate system.
Thus, the entire phase space of the stadium is explored.     

The results for the average and the standard deviation of the intensities 
using Birkhoff coordinates are shown in Figs.~(\ref{avg_intensity}) 
and (\ref{rms_intensity}), respectively.  The average is flat except for peaks 
associated with the two symmetry lines of the stadium.  The eigenstates used 
here were even-even states, so there is twice the intensity along the two symmetry 
lines that bisect the end caps and the straight edges.   The standard deviation has 
two large peaks centered around the bouncing ball orbits.  The rest of the 
figure is relatively flat with a few small bumps.  Random matrix theory 
would predict this to be a flat figure with small oscillations.  
The marginal stability of the bouncing ball orbits can be seen but no other 
feature of the stadium, except for the horizontal bounce, is picked out by 
looking at the intensities.  Fig.~(\ref{intensity_variance}) shows the 
$\hbar$-scaling of the root mean square for the intensities where the wave packet 
is placed on various periodic orbits.  The theory from Section~\ref{semi_dynamics}
predicts a smaller scaling than the numerical results of the stadium.

\begin{figure}
  \epsfig{file=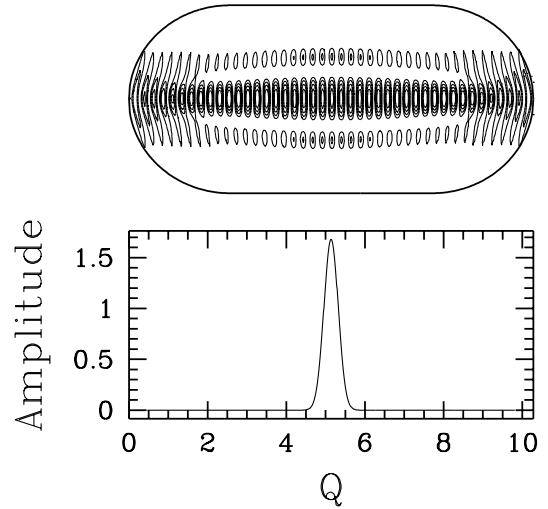, width=8.2cm}
  \caption{The lower figure is Gaussian wave packet on the boundary. 
  The upper figure corresponds to the wave packet in the interior of the stadium.}
  \label{gaussian}
\end{figure}

\begin{figure}
  \epsfig{file=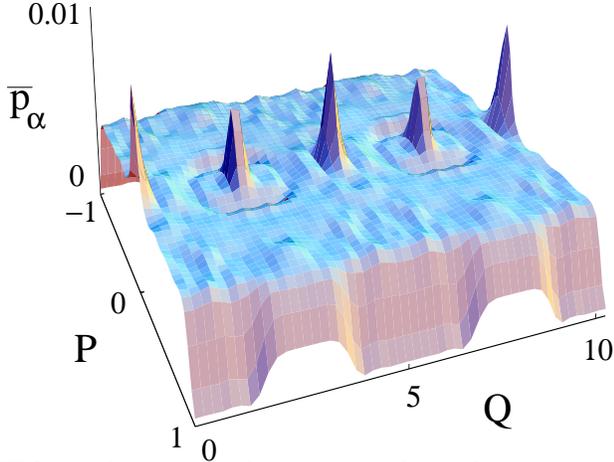, width=8.2cm}
  \caption{Average overlap intensity for a Gaussian wave packet defined in
  the Birkhoff coordinates of the stadium.}
  \label{avg_intensity}
\end{figure}

The heights of the bouncing ball peaks can be approximated by quantizing
the rectangular region of the stadium.  The intensities obtained from
this calculation are weighted by the ratio of the density of the bouncing
ball states~\cite{sieber} to the total density of states.  The Gaussian 
wave packet is placed in the center of the straight edge and the middle 
energy regime is used.  The results of this approximation are 80.6 and 320.5 
for the average intensity and rms intensity, respectively, compared to 80.7 
and 386.3 for the numerical calculations of the stadium.

\begin{figure}
  \epsfig{file=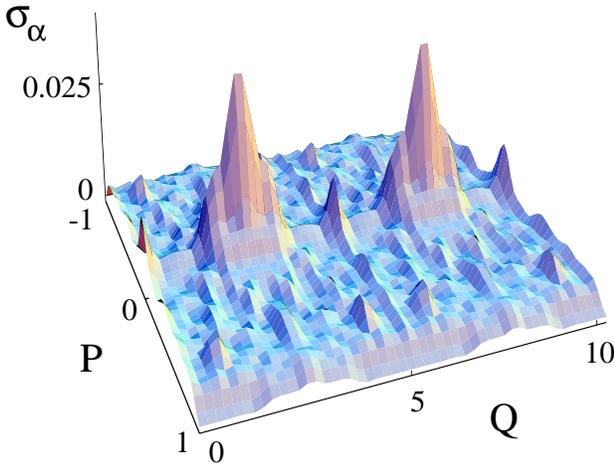, width=8.2cm}
  \caption{Root mean square of the overlap intensity for a Gaussian wave packet 
  defined in the Birkhoff coordinates of the stadium.}
  \label{rms_intensity}
\end{figure}

Random matrix theory suggests that the correlation coefficient for a
generic chaotic system should result in zero.  On the other hand, 
using the correlation coefficient for the stadium in Birkhoff 
coordinates, we found that some of the states gave nonzero 
correlations, Fig.~(\ref{overlap_coeff}).  In fact, large
correlations are found for nearly all the states in the stadium
billiard which means that there exists phase space localization
for most of the states.  The large positive values of the correlation 
coefficient in the center of the figure again correspond to the bouncing
ball states.  Classically, this area of phase space is difficult
to enter and leave.  Hence, the localization is expected
to be stronger for this area of phase space.  The area beneath the
peaks is several standard deviations $(N^{-1/2} = (114)^{-1/2} \approx 0.09)$
away from  zero as predicted by random matrix theory.  Thus, phase space 
localization is also occurring in this region.  The point exactly 
in between the peaks is the point in phase space associated with the 
horizontal bounce.  The series of smaller peaks leading up to the large 
peaks are the gateways into the vertical bouncing ball area.  Fig.~(\ref{trajectory}) 
is a plot of the orbits corresponding to these peaks.  They are
periodic orbits which only strike the endcaps twice and become almost
vertical.  Orbits must pass through these regions in order to enter or 
exit the vertical bounce states.

\begin{figure}
  \epsfig{file=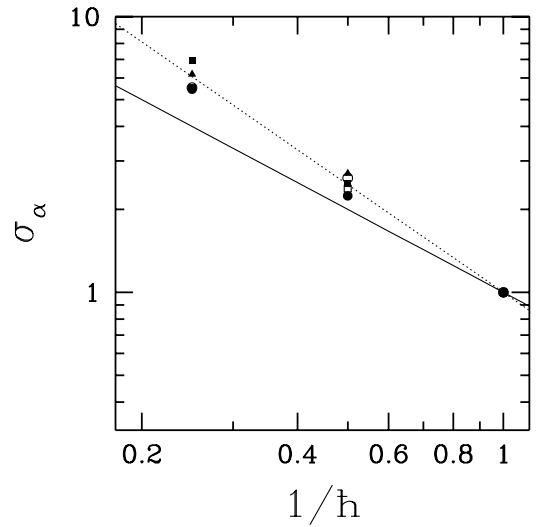, width=8.2cm}
  \caption{Root mean square of the overlap intensities as a function of $1/\hbar$.  
  The wave packet was placed on the horizontal (solid circles), V (solid squares), 
  diamond (solid triangles), rectangle (open circles) and bow tie (open squares) 
  orbits.  The solid line is the theorectial value of $\hbar^1$ from 
  Section~\ref{semi_dynamics} and the dotted line is the best fit of the 
  stadium results which is $\hbar^{1.3}$.  The intensities have been rescaled 
  at $\hbar = 1$ so that they occupy the same area of the plot.}
  \label{intensity_variance}
\end{figure}

\begin{figure}
  \epsfig{file=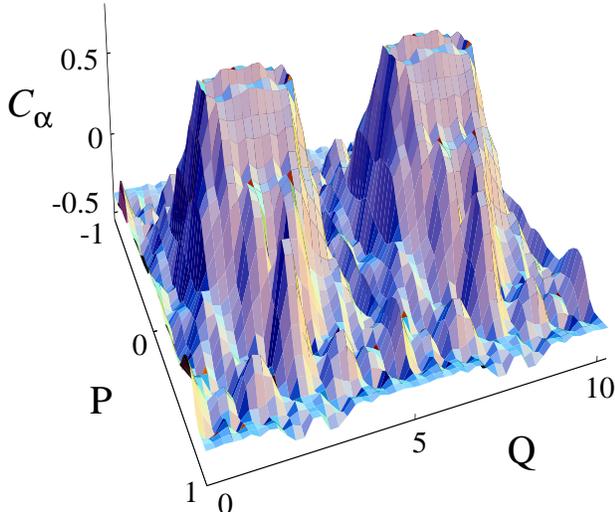,width=8.2cm}
  \caption{Overlap correlation coefficient for the stadium using
  Birkhoff coordinates.  The energy range of the averaging is
  2200-2600 where $\hbar = m = 1$.}
  \label{overlap_coeff}
\end{figure}

As the energy of the system is increased (i.~e.~$\hbar$ is decreased), 
the results of the correlation function remain qualitatively the same, 
Fig.~(\ref{overlap_coeff2}).  All the peaks and valleys stay in the same place.  
The numerical results of the overlap correlation coefficient fluctuate depending
upon the area of phase space being considered.  This is consistent with the 
the semiclassical theory in Section~\ref{semi_dynamics}.  More details of the
system are explored as $\hbar$ is decreased, since the phase space is divided into
finer areas.  Thus, more detailed information about the phase space localization 
of the system is observed in the overlap correlation coefficient at smaller values 
of $\hbar$.

\begin{figure}
  \epsfig{file=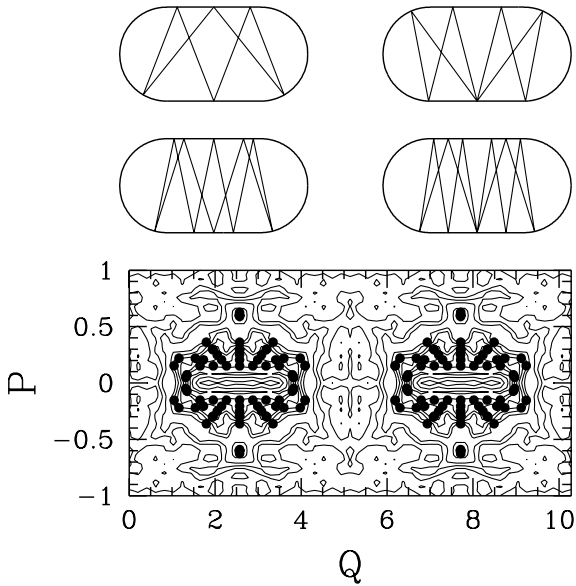, width=8.2cm}
  \caption{Trajectories corresponding to the peaks leading up to
  the bouncing ball orbits.  The lower figure is a contour plot of 
  Fig.~(\ref{overlap_coeff2}).  The solid circles correspond to the 
  bounce points of the trajectories.  Geometric and time-reversal 
  symmetries were also included.}
  \label{trajectory}
\end{figure}

\end{section}

\begin{section}{Conclusions}
\label{conclusions}

We have shown that intensity weighted level velocities are a good
measure of the localization properties for chaotic systems.  They
are far more sensitive to localization than similarly weighted level
curvatures (which are closely related to level statistics).  Thus,
a system can be RMT-like, yet the eigenstates are not behaving
ergodically (as RMT predicts).

The stadium eigenstates show a great deal of localization.  Not only
are the vertical bouncing ball orbits predicted by the measure, but
also other orbits.  The overlap correlation coefficient is very parameter 
dependent.  Choosing a different parameter to vary would highlight other 
sets of orbits depending on how strong the perturbation effects those orbits.   
The degree of localization can be predicted by
the return dynamics.  In a chaotic system, all the return dynamics
can be organized by the homoclinic orbits.  The manner in which a 
chaotic system's eigenstates approach ergodicity as $\hbar \rightarrow 0$
will depend on a new time scale, i.e.~that required for the homoclinic
excursions to explore the available phase space fully.  

\begin{figure}
  \epsfig{file=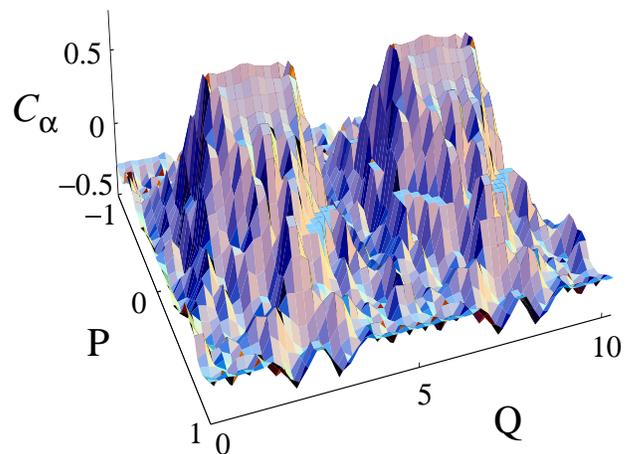, width=8.2cm}
  \caption{The same as Fig.~(\ref{overlap_coeff}) except for a higher energy range
  of 9200-10000 where $\hbar = m = 1$.  Note the finer structure of the various
  peaks.}
  \label{overlap_coeff2}
\end{figure}

Parametrically varied data exist that can be analyzed in this way.
In the Coulomb-blockade conductance data to the extent that the resonance 
energy variations are related to a single particle level velocity
(minus a constant charging energy and absent residual interaction
effects) should show correlations.  We mention also that the 
microwave cavity data can be studied with even more flexibility since they
have measured the eigenstates and can therefore meticulously
study a wide range of $|\alpha \rangle$ to get a complete picture of
the eigenstate localization properties.

Finally, this analysis could be applied in a very fruitful way to
near-integrable and mixed phase space systems.  In these cases, 
standard random matrix theory would not give the zero$^{th}$ order
statistical expectation, but the localization would still be 
determined by the return dynamics in the semiclassical approximation.

\acknowledgements

We gratefully acknowledge important discussions with B.~Watkins and T.~Nagano
and support from the National Science Foundation under Grant No. NSF-PHY-9800106 
and the Office of Naval Research under Grant No. N00014-98-1-0079.  

\end{section}

\appendix

\begin{section}{Gaussian Integration}
\label{A}

Inserting Eq.~(\ref{green}) into Eq.~(\ref{strength_osc}), the strength
function involves two $N$-dimensional integrals where $N$ is the system's 
number of degrees of freedom,
\begin{eqnarray}
  S_{\alpha,osc}(E) &=& {-1 \over \pi} {\mathrm Im}
    {1 \over i \hbar (2 \pi i \hbar)^{(d - 1)/2}} 
    \left( {1 \over \pi \sigma^2} \right)^{d/2} \nonumber \\
  && \times \int \sum_{j} |D_s|^{1/2}
    \exp \{-i{\bf p}_\alpha \cdot ({\bf q} - {\bf q}^\prime)/\hbar \nonumber \\
  && - ({\bf q} - {\bf q}_\alpha)^2 / 2\sigma^2
    - ({\bf q}^\prime - {\bf q}_\alpha)^2 / 2\sigma^2 \nonumber \\
  && + iS_j({\bf q}, {\bf q}^\prime;E)/\hbar - i \nu_j^\prime\pi/2 \} 
    d{\bf q} d{\bf q}^\prime
\end{eqnarray}
To evaluate the integrals over ${\bf q}$ and ${\bf q}^\prime$, the action is 
quadratically expanded about the points ${\bf q}_f$ and ${\bf q}_i$,
\begin{eqnarray}
  S_j({\bf q}, {\bf q}^\prime;E) &=& S_j({\bf q}_f, {\bf q}_i;E)
    + {\bf p}_f \cdot ({\bf q} - {\bf q}_f)
    - {\bf p}_i \cdot ({\bf q}^\prime - {\bf q}_i)
    \nonumber \\
  &+& {1 \over 2} \sum^{N}_{i,k} \left[ 
    \left( {\partial p^{(i)}_f 
    \over \partial q^{(k)}} \right)_{{\bf q}_f}
    (q^{(i)} - q^{(i)}_f) (q^{(k)} - q^{(k)}_f) \right. \nonumber \\
  &-& \left. \left( {\partial p^{(i)}_i 
    \over \partial q^{\prime (k)}} \right)_{{\bf q}_i}
    (q^{\prime (i)} - q^{(i)}_i) (q^{\prime (k)} - q^{(k)}_i) \right. \\
  &+& \left. 2 \left( {\partial p^{(i)}_f 
    \over \partial q^{\prime (k)}} \right)_{{\bf q}_i}
    (q^{(i)} - q^{(i)}_f) (q^{\prime (k)} - q^{(k)}_i) \right] \nonumber
\end{eqnarray}
It is useful to define the vector
\begin{equation}
  {\bf z} = (z_1, \dots, z_{N}, z^\prime_1, \dots, z^\prime_{N})
\end{equation}
where
\begin{eqnarray}
  z_i &=& (q^{(i)} - q^{(i)}_f)/\sigma \nonumber \\
  z^\prime_i &=& (q^{\prime (i)} - q^{(i)}_i)/\sigma
\end{eqnarray}
Thus, the integrals become
\begin{eqnarray}
  S_{\alpha,osc}(E) &=& {-1 \over \pi} {\mathrm Im}
    {1 \over i \hbar (2 \pi i \hbar)^{(d - 1)/2}} 
    \left( {1 \over \pi \sigma^2} \right)^{d/2} \nonumber \\
  && \times \int \sum_{j} |D_s|^{1/2} \sigma^{2d} \nonumber \\ 
  && \times \exp \{-{\bf z} \cdot {\bf A} \cdot {\bf z}
    - {\bf b} \cdot {\bf z} + c\} d{\bf z}
\end{eqnarray}
where ${\bf A}$ is composed of four $N$-dimensional matrices
\begin{equation}
  \label{A_matrix}
  {\bf A} = \left( 
  \begin{array}{cc}
    {\bf A}^{11} & {\bf A}^{12} \\
    {\bf A}^{21} & {\bf A}^{22} \\
  \end{array}
  \right)
\end{equation}
and
\begin{eqnarray}
  {\bf b} &=& \left( i \delta p^{(1)}_f - \delta q^{(1)}_f,
    \dots,  i \delta p^{(N)}_f - \delta q^{(N)}_f, \right. \nonumber \\
  && \left. i \delta p^{(1)}_i - \delta q^{(1)}_i,
    \dots,  i \delta p^{(N)}_i - \delta q^{(N)}_i \right)
\end{eqnarray}
with
\begin{eqnarray}
  \label{delta_qp}
  \delta  p^{(i)}_f &=& (p^{(i)}_\alpha - p^{(i)}_f)\sigma / \hbar \nonumber \\
  \delta  p^{(i)}_i &=& (p^{(i)}_\alpha - p^{(i)}_i)\sigma / \hbar \nonumber \\
  \delta  q^{(i)}_f &=& (q^{(i)}_\alpha - q^{(i)}_f) / \sigma  \nonumber \\
  \delta  q^{(i)}_i &=& (q^{(i)}_\alpha - q^{(i)}_i) / \sigma 
\end{eqnarray}
and
\begin{eqnarray}
  c &=& {i \over \hbar} S_j({\bf q}_f, {\bf q}_i; E) 
    - {i \over \hbar} {\bf p}_\alpha \cdot ({\bf q}_f - {\bf q}_i) \nonumber \\
  && - {({\bf q}_f - {\bf q}_\alpha)^2 \over 2 \sigma^2}
    - {({\bf q}_i - {\bf q}_\alpha)^2 \over 2 \sigma^2}
    - {i \nu_j^\prime \pi \over 2}
\end{eqnarray}
The matrix ${\bf A}$ can be expressed in terms of the stability matrix,
${\bf M}$, where ${\bf M}$ has the same form as Eq.~(\ref{A_matrix})
\begin{equation}
  \left( 
  \begin{array}{c}
    {\bf p} \\ {\bf q}	
  \end{array}
  \right) = {\bf M} \left(
  \begin{array}{c}
    {\bf p}^\prime \\ {\bf q}^\prime	
  \end{array}
  \right)
\end{equation}
Thus,
\begin{eqnarray}
  {\bf A}^{11}_{ab} &=& {\delta_{a,b} \over 2} - {i \sigma^2 \over 2 \hbar} 
    \left( {\partial p^{(a)}_f \over \partial q^{(b)}} \right)_{{\bf q}_f} 
    \nonumber \\
  &=& {\delta_{a,b} \over 2} - {i \sigma^2 \over 2 \hbar}
    {\sum^{N}_{i} m_{a,i} {\mathrm cof} \left( {\bf M}^{21}_{b,i} \right) \over
    \det {\bf M}^{21}} \nonumber \\
  &=& {{\bf I} \over 2} - {i \sigma^2 \over 2 \hbar}
    {\bf M}^{11} ({\bf M}^{21})^{-1} \nonumber \\
  {\bf A}^{12}_{ab} &=& -{i \sigma^2 \over 2 \hbar} 
    \left( {\partial p^{(a)}_f \over \partial q^{\prime (b)}} \right)_{{\bf q}_i}
    \nonumber \\
  &=& {i \sigma^2 \over 2 \hbar} {{\mathrm cof} \left( {\bf M}^{21}_{b,a} \right)
    \over \det {\bf M}^{21}} \nonumber \\
  &=& {i \sigma^2 \over 2 \hbar} (({\bf M}^{21})^{-1})^T \nonumber \\
  {\bf A}^{21}_{ab} &=&  {i \sigma^2 \over 2 \hbar} 
    \left( {\partial p^{(a)}_i \over \partial q^{(b)}} \right)_{{\bf q}_f}
    \nonumber \\
  &=& {i \sigma^2 \over 2 \hbar} {{\mathrm cof} \left( {\bf M}^{21}_{a,b} \right)
    \over \det {\bf M}^{21}} \nonumber \\
  &=& {i \sigma^2 \over 2 \hbar} ({\bf M}^{21})^{-1} \nonumber \\
  {\bf A}^{22}_{ab} &=&{\delta_{a,b} \over 2} + {i \sigma^2 \over 2 \hbar} 
    \left( {\partial p^{(a)}_i 
    \over \partial q^{\prime (b)}} \right)_{{\bf q}_i} \nonumber \\ 
  &=& {\delta_{a,b} \over 2} - {i \sigma^2 \over 2 \hbar}
    {\sum^{N}_i m_{i + N,b + N} 
    {\mathrm cof} \left( {\bf M}^{21}_{i,a} \right) \over \det {\bf M}^{21}}
    \nonumber \\
  &=& {{\bf I} \over 2} - {i \sigma^2 \over 2 \hbar}
    ({\bf M}^{21})^{-1} {\bf M}^{22}
\end{eqnarray}
where $m_{i,k}$ are elements of the stability matrix and 
${\mathrm cof} \left( {\bf M}^{21}_{ik} \right)$ is the signed minor of
${\bf M}^{21}_{ik}$.  $D_s$ is a determinate involving second derivatives of 
the actions,
\begin{eqnarray}
  D_s = \left|
  \begin{array}{cc}
    {\partial^2 S \over \partial {\bf q} \partial {\bf q}^{\prime}} 
    &  {\partial^2 S \over \partial {\bf q} \partial E} \\
    {\partial^2 S \over \partial E \partial {\bf q}^{\prime}}
    & {\partial^2 S \over \partial E^2} \\
  \end{array}
  \right|
  &=& {1 \over |\dot{q}^{(N)}| |\dot{q}^{\prime (N)}|} 
    \left| {- \partial^2 S 
    \over \partial \tilde{\bf q} \partial \tilde{\bf q}^\prime} \right|
    \nonumber \\
  &=& {1 \over |\dot{q}^{(N)}| |\dot{q}^{\prime (N)}|}
    \left( {2 \hbar \over i \sigma^2} \right)^{(d - 1)} \nonumber \\ 
  && \times \left| \tilde{{\bf A}}^{21} \right|
\end{eqnarray}
The tildes in the determinants in the above equation are used to exclude the 
$N$th coordinate.  To obtain the above result $q^{(N)}$ and $q^{\prime (N)}$ 
are chosen to be locally oriented along the trajectory where the dots indicate
time derivatives.  Since ${\bf A}$ is a symmetric matrix, the result for a general 
Gaussian integral is used and, hence, the strength function becomes
\begin{eqnarray}
  S_{\alpha,osc}(E) &=& {-1 \over \pi} {\mathrm Im}
    {1 \over i \hbar (2 \pi i \hbar)^{(d - 1)/2}} 
    \left( {\sigma^2 \over \pi} \right)^{d/2} 
    \left( {2 \hbar \over i \sigma^2} \right)^{(d - 1)/2} \nonumber \\
  && \sum_{j} \left( {\pi^{2d} \over \det{\bf A}} \right)^{1/2}
    \left( { 1 \over |\dot{q}^{(N)}| |\dot{q}^{\prime (N)}|} \right)^{1/2} 
    \left| \tilde{{\bf A}}^{21} \right|^{1/2} \nonumber \\
  && \times \exp \left\{{1 \over 4}{\bf b} \cdot {\bf A}^{-1} \cdot {\bf b} + c \right\} 
    \nonumber \\
  &=& {\sigma \over \pi^{1/2} \hbar} {\mathrm Re} \sum_j
    \left( {\det \tilde{{\bf A}}^{21} \over \det{\bf A}} \right)^{1/2}
    \left( { 1 \over |\dot{q}^{(N)}| |\dot{q}^{\prime (N)}|} \right)^{1/2}
    \nonumber \\
  && \times \exp \left\{{1 \over 4}{\bf b} \cdot {\bf A}^{-1} \cdot {\bf b} + c \right\}
\end{eqnarray} 
where the time derivatives are evaluated at the saddle points.

\end{section}

\begin{section}{Sum Rule for the Strength Function}
\label{B}

The determinant of the $(2N \times 2N)$ matrix ${\bf A}$,
\begin{equation}
  \det {\bf A} = \det \left( 
  \begin{array}{cc}
    {{\bf I} \over 2} - {i \sigma^2 \over 2 \hbar} {\bf M}^{11} ({\bf M}^{21})^{-1}
      & {i \sigma^2 \over 2 \hbar} (({\bf M}^{21})^{-1})^T \\
    {i \sigma^2 \over 2 \hbar} ({\bf M}^{21})^{-1} 
      & {{\bf I} \over 2} - {i \sigma^2 \over 2 \hbar} 
      ({\bf M}^{21})^{-1} {\bf M}^{22}\\
  \end{array}
  \right)
\end{equation}
can be reduced to determinants of $(N \times N)$ matrices by using the relation
$(({\bf M}^{21})^{-1})^T 
= -{\bf M}^{12} + {\bf M}^{11} ({\bf M}^{21})^{-1} {\bf M}^{22}$ 
and some row and column manipulations~\cite{gutzwiller}, so that
\begin{eqnarray}
  \label{det_A}
  \det {\bf A} &=& \det \left({-i \sigma^2 \over 4 \hbar} 
    \left[{\bf M}^{11} + {\bf M}^{22} \right. \right. \nonumber \\ 
  && \left. \left. + i\left( {\hbar \over \sigma^2} {\bf M}^{21} 
    - {\sigma^2 \over \hbar} {\bf M}^{12} \right)\right]\right) / \det({\bf M}^{21})
\end{eqnarray}
The coordinates parallel to the trajectory do not mix with the transverse
coordinates, since a point on an orbit will remain on that particular orbit.
Thus, the $N$th rows and columns of the individual matrices in the above
expression are zero except for the $(N,N)$ elements.  

It is convenient to re-express the submatrices of the stability matrix in terms of 
the Lyapunov exponents.  Let $\{\lambda_i\}$ be the set of Lyapunov exponents 
whose real part is positive ordered such that 
$\lambda_1 > \lambda_2 > \cdots > \lambda_{N - 1}$.
The Lyapunov exponents along the parallel coordinate are zero and we will only 
work with the reduced $(2(N-1) \times 2(N-1))$ stability matrix in what follows.
Let ${\bf \Lambda}$ be the diagonal matrix of the 
eigenvalues of the reduced stability matrix,
\begin{equation}
  {\bf \Lambda} = \left( 
  \begin{array}{cccccc}
    e^{\lambda_1 t} & \cdots & 0 & 0 & \cdots & 0 \\
    \vdots &  & \vdots & \vdots & & \vdots \\	
    0 & \cdots & e^{\lambda_{N - 1} t} & 0 & \cdots & 0 \\
    0 & \cdots & 0 & e^{-\lambda_1 t} & \cdots & 0 \\
    \vdots &  & \vdots & \vdots & & \vdots \\	
    0 & \cdots & 0 & 0 & \cdots & e^{-\lambda_{N - 1} t} \\
  \end{array}
  \right)
\end{equation}
Thus, by a similarity transform the reduced stability matrix be can written 
in terms of the Lyapunov exponents, i.e.
\begin{equation}
  {\bf M} = {\bf L} {\bf \Lambda} {\bf L}^{-1}
\end{equation}
Hence, each of the elements of the stability matrix can be written as
\begin{equation}
  \label{matrixelements}
  m_{ij} = \sum_k^N a_{ij}^{(k)} e^{\lambda_k t} + b_{ij}^{(k)} e^{-\lambda_k t}
\end{equation}
where $a_{ij}^{(k)}$ and $b_{ij}^{(k)}$ are linear combinations of the elements of
the ${\bf L}$ and ${\bf L}^{-1}$ matrices.  Because in general chaotic systems
$\lambda \gg 0$, the $b_{ij}^{(k)}$'s may be omitted without seriously effecting
the above sum.  All the determinants including the numerator and denominator of 
$\det{\bf A}$ as well as $\det \tilde{{\bf A}}^{21}$, thus, will involve products 
of Eq.~(\ref{matrixelements}).  The homoclinic orbits in the sum begin and end at 
the intersections of the stable and unstable manifolds near the Gaussian centroid.
Since neither manifolds may cross themselves, then in the vicinity
of the Gaussian centroid the branches of each manifold are nearly parallel
to themselves.  Thus, to an excellent approximation, the same similarity 
transformation will diagonalize the stability matrix for each individual orbit, 
regardless of the period.  Consequently, the elements of 
${\bf L}$ and ${\bf L}^{-1}$ are period independent.

Connections can be made between the determinants and the Kolmogorov-Sinai entropy.
The Kolmogorov-Sinai entropy, $h_{KS}$, of a system can be expressed using Pesin's 
Theorem as the sum of the Lyapunov exponents with positive real part,
\begin{equation}
  h_{KS} = \sum_i^{N - 1} \lambda_i
\end{equation}
If there is no mixing between the different coordinates, then the individual
matrices ${\bf M}^{11}$, ${\bf M}^{12}$, ${\bf M}^{21}$ and ${\bf M}^{22}$
are diagonal.  Thus, each matrix element depends only upon one Lyapunov exponent
and the determinants are proportional to 
$\exp (-h_{KS} t)$.  This is the case for two dimensional systems where the
parallel and perpendicular coordinates in the stability matrix separate as mentioned
above.  Hence, we have
\begin{equation}
  \left| {\det \tilde{{\bf A}}^{21} \over \det{\bf A}} \right|
  \propto \exp (-h_{KS} t)
\end{equation}

Unlike the periodic orbits, the homoclinic sum is over segments of the orbits.
The number of homoclinic points will proliferate exponentially 
at the same rate as the fixed points in the neighborhood which is 
proportional to $\exp(h_T T)$ where $h_T$ is the topological entropy.
This is demonstrated by examining the partitioning of the phase space
mentioned in Sect.~\ref{overlapintensities} which has exponential growth.
The partitioning reflects the symbolic dynamics of the system.  The symbolic
code uniquely describes each orbit so that amount of code (partitions)
cannot grow faster than the number of periodic points, since each code (partition) 
cannot represent more than one periodic point.

Finally, the sum rule is obtained by setting the topological entropy and 
Kolmogorov-Sinai entropy equal to each other. Then, for the special case 
of no mixing in the stability matrix as mentioned above the combination 
of the amplitudes and the number of orbits yields
\begin{equation}
  \label{homosumrule}
  \sum_j \left| {\det \tilde{{\bf A}}^{21} \over \det{\bf A}} \right| \cdots
  \rightarrow \int dT \cdots
\end{equation}

\end{section}

\end{document}